\documentclass[twocolumn,aps,pra,superscriptaddress]{revtex4-2}
\usepackage{amsmath,amsfonts,amssymb}
\usepackage{braket}
\usepackage{ulem} 
\usepackage[dvipsnames]{xcolor}
\usepackage{comment}
\usepackage{graphicx}
\usepackage{xcolor}
\usepackage{dsfont}
\usepackage{hyperref}
\usepackage{tikz}
\usepackage{subcaption}
\usepackage{siunitx}

\newcommand{\ex}[1]{\langle #1\rangle}

\begin{document}

\title{Measurement-based Feedback Control of a Quantum System in a Harmonic Potential}
\author{Amy Rouillard}
\affiliation{School of Chemistry and Physics, University of KwaZulu-Natal, Private Bag X54001, Durban 4000, South Africa}
\author{Anirudh Reddy}
\affiliation{School of Chemistry and Physics, University of KwaZulu-Natal, Private Bag X54001, Durban 4000, South Africa}
\author{Humairah Bassa}
\affiliation{School of Chemistry and Physics, University of KwaZulu-Natal, Private Bag X54001, Durban 4000, South Africa}
\author{Shamik Maharaj}
\affiliation{School of Chemistry and Physics, University of KwaZulu-Natal, Private Bag X54001, Durban 4000, South Africa}
\author{Lajos Di\'{o}si}
\affiliation{Wigner Research Centre for Physics, H-1525 Budapest, Hungary}
\affiliation{Department of Physics of Complex Systems, E\"{o}tv\"{o}s Lor\'{a}nd University, P\'{a}zm\'{a}ny P\'{e}ter stny. 1/A, H-1117 Budapest, Hungary}
\author{Thomas Konrad}
\email{konradt@ukzn.ac.za}
\affiliation{School of Chemistry and Physics, University of KwaZulu-Natal, Private Bag X54001, Durban 4000, South Africa}
\affiliation{National Institute of Theoretical and Computational Sciences (NITheCS), KwaZulu-Natal, South Africa}

\begin{abstract}
We present a formulation of measurement-based feedback control of a single quantum particle in one spatial dimension. An arbitrary linear combination of the position and momentum of the particle is continuously monitored, and feedback proportional to the measured signal is used to control the system. We derive a feedback master equation and discuss a general approach to computing the steady-state solutions for arbitrary potentials. For a quantum harmonic oscillator or a free particle, we show that it is possible to cool and confine the system using feedback that simultaneously damps the measured observable and its conjugate momentum, as well as compensates for noise introduced by the measurement. In addition, we demonstrate that appropriate feedback adds a quadratic term in the measured observable to the Hamiltonian of the system. For the particular case of the harmonic potential, we describe the exact dynamics of the system that can be cooled to the ground state. Moreover, we provide an argument for the possibility to cool systems with arbitrary potentials provided that the measurement is strong enough to localise the particle on an interval smaller than the characteristic length scale of the potential. 
\end{abstract}

\maketitle

\section{Introduction}

Recent developments in the fields of quantum computing and quantum meteorology have resulted in a demand for protocols to control individual quantum systems\cite{DAlessandro,Nielsen2011,d2021introduction}. One paradigm of quantum control is measurement-based feedback control in which the dynamics of a system are altered as information received from continuous measurements is fed back into the system. For example, in order to impede the movement of a particle a force proportional to the measured velocity can be applied in the opposite direction to that of the motion. Although a measurement inevitably changes the state of a quantum system, feedback control can be achieved with weak or unsharp measurements \cite{WisemanBook}, which allow to gain information about a system without much disturbance.

In this paper we present the theory that describes the dynamics of a system that is controlled by instantaneously feeding back the measurement signal at each moment, which corresponds to the Markovian limit. This is done by introducing a feedback Hamiltonian which depends on the stochastic measurement signal. The possibility of continuous feedback control was studied theoretically by Wiseman and Milburn\cite{WisemanBook}  as well as experimentally in \cite{bushev2006feedback, steixner2005quantum, setter2018real}. In particular, there are studies of the cooling of a single trapped ion \cite{bushev2006feedback, steixner2005quantum}, the cooling of an optically levitated nano-particle \cite{setter2018real}, and the cooling of trapped atoms probed by off-resonant light \cite{leibfried2003quantum}.

Feedback presupposes a measurement step which necessarily introduces noise into the process even if all classical noise sources are eliminated. In our theory, this is manifest as a diffusion-like term in the master equation. The beauty of direct quantum feedback is that it might be used to cancel the noise which drives the mean values of the dynamical variables as shown in Section~\ref{section:Analysing_feedback}. This is possible because the noise in the measurement signal is the same noise that drives the system. The result is that the steady-state of the feedback master equation can have the same variances as the conditioned states since all of the fluctuations in the mean values are overcome. %

Here, we use the method developed by Diosi in \cite{diosi1988localized} to express the state change in terms of generators and in the "co-moving" frame, that is the reference frame in which the expectation values of both position and momentum are zero.  We show that this method can greatly reduce the complexity of computations without additional approximation, as well as give insight into the type of feedback necessary to achieve the desired control. 

In order to illustrate the power of this method we consider the case that the measured and feedback observables are an arbitrary linear combination of the position and momentum. We derive a stochastic feedback master equation in Section~\ref{section:master_equation} and compute the corresponding steady-state in Section~\ref{sec:stat_har}. In addition, we show that appropriate feedback can be used to compensate a part of the Hamiltonian of the system (Section~\ref{section:Analysing_feedback}).  In addition, for a harmonic potential, we determine how feedback can confine the particle to the centre of phase space in the laboratory frame, thereby reducing the energy of the particle. This phenomenon is also discussed for other potentials.

\section{The Feedback Master Equation}\label{section:master_equation}

Our feedback scheme relies on the use of information obtained from continuous observation to manipulate the motion of the system, e.g. to cool or confine it. 
Assuming the measured observable is associated with the operator $\hat{M}$, the increment of the readout of the continuous measurement at a certain instance is
\begin{equation}\label{eq:dM}
  d\mathcal{M} = \langle \hat{M} \rangle dt + \tfrac{1}{\sqrt{\gamma}} dW
\end{equation}
where  $\gamma$ is the measurement strength which determines the rate at which information is extracted from the system. 
The first term is the expectation value of the observable, $\ex{\hat{M}} \equiv\mathrm{Tr}[\hat{M} \rho]$, defined with respect to the density operator $\rho$ in the selective regime which represents the current state of the system for a given measurement record. The second term represents a contribution of Gaussian white noise ($dW$), modelling the deviation of the measurement result from the expectation value of the observable by a random Wiener process.

According to continuous measurement theory \cite{WisemanBook, jacobs2014quantum}, the state change of the system, due to the measurement of the observable after a time step $dt$, is given by the Ito stochastic differential equation 
\begin{align}
  d\rho = -&\tfrac{i}{\hbar}[\hat{H},\rho]dt - \tfrac{\gamma}{8}\left[\hat{M},\left[\hat{M},\rho\right]\right]dt \nonumber \\
  &+ \tfrac{\sqrt{\gamma}}{2}\left\{\hat{M} - \langle \hat{M}\rangle,\rho\right\}\mathrm{d}W.\label{eq:ito_dif}
\end{align}
The first term corresponds to the unitary evolution related to the Hamiltonian $\hat{H}$ of the system. The second term refers to the dissipative dynamics due to measurement  and tends to diagonalize the density operator in the eigenbasis of the measured observable, $\hat{M}$.  The final term is proportional to the Wiener increment, $dW$, and indicates an update of the observer's knowledge about the system according to a certain measurement result, $d\mathcal{M}$. The Ito differential $dW$ satisfies the following algebra:
\begin{align}
    \langle dW \rangle_{st} &= 0, \nonumber \\
    dWdW &= dt, \nonumber \\
    dW^n & =0\text{ for }n>2\,.\label{eq:Ito_rules}
\end{align}
The Wiener noise interment $dW$ at a given time $t$, points to the future and is thus statistically independent of the state $\rho$ and the measurement signal~\eqref{eq:dM} at time $t$. 

A simple Markovian feedback can be achieved by amplifying and feeding back the measurement signal, via the Hamiltonian~\cite{combes2011quantum}
\begin{equation}
    \hat{H}_{\mathrm{fb}}(t) = \hat{F}\frac{d\mathcal{M}}{\delta t}\,,
\end{equation}
where $\hat{F}$ represents an hermitian operator, and $\delta t$ is a sufficiently small time interval during which the feedback is implemented by means of the unitary time evolution
\begin{equation}\label{eq:unitary_operator}
  U = \exp\left[-\frac{i}{\hbar} \hat{F} d\mathcal{M} \right]\,.
\end{equation}
Although Markovian feedback is not as powerful as Bayesian feedback \cite{audretsch2007monitoring}, which uses the best estimate of the system state, it is theoretically and experimentally simpler to implement~\cite{combes2011quantum,wiseman2002bayesian}. Realistic feedback is not Markovian, but has some time delay. For the Markovian approximation to be satisfied, the time delay between measurement and feedback should be small compared to the time scale on which the system evolves. We point out that it is possible to implement feedback without delay by incorporating it into the interaction with a probe system, that also realises the measurement \cite{konrad2021robust}.  

A system undergoing continuous observation and Hamiltonian evolution changes as $\rho \rightarrow \rho + d\rho$, according to Eq.~\eqref{eq:ito_dif}. Adding feedback, described by the unitary operator given in Eq.~\eqref{eq:unitary_operator}, causes the system to evolve as $\rho \rightarrow U(\rho+d\rho)U^\dagger$. The resulting  stochastic master equation for the density operator of the system is thus given by
\begin{align}\label{eq:stochastic_master}
  d\rho =& -\frac{i}{\hbar}\left[\hat{H},\rho\right]dt - \frac{\gamma}{8}\left[\hat{M},\left[\hat{M},\rho\right]\right]dt  \nonumber \\
  &-\frac{1}{2\hbar^2 \gamma}\left[\hat{F},\left[\hat{F},\rho\right]\right]dt -\frac{i}{2\hbar }\left[\hat{F}, \left\{\hat{M},\rho\right\}\right]dt   \nonumber \\
  &+ \frac{\sqrt{\gamma}}{2}\left\{\hat{M} - \langle \hat{M} \rangle,\rho\right\}dW - \frac{i}{\hbar \sqrt{\gamma}}\left[\hat{F},\rho\right]dW\,,
\end{align}
and has a Lindblad structure. The complete derivation is given in Appendix~\ref{ap:ito}. This expression also appears in \cite{WisemanBook, tilloy2016sourcing}. The first, second and fifth terms  of this equation describes the free evolution and measurement back-action, while the remaining terms include the effects of the feedback loop. The fourth term causes the expected damping while the third term leads to diffusion in the observable conjugate to the generator of feedback $\hat{F}$. The diffusion occurs as a result of the noise in the measurement signal, which is also amplified and fed back to the system.


\section{Determining the stationary state using the co-moving frame}\label{section:steady_state}
Here we follow the method of Diosi \cite{diosi1988localized} in order to derive the stationary-state solution. Accordingly, we use the coordinate free representation and assume the state of the system to be a pure state $\ket{\psi}$. We can thus write the feedback equation for the state vector in the equivalent form 
\begin{align}
     d\ket{\psi} = &\Big\{ \Big[-\tfrac{i}{\hbar}\hat{H} - \tfrac{\gamma}{8}\left(\hat{M} - \ex{\hat{M}}\right)^2 \nonumber\\ & \quad -\tfrac{1}{2\hbar^2 \gamma}\hat{F}^{2}
      -\tfrac{i}{2\hbar}\hat{F}\left(\hat{M}+ \ex{\hat{M}}\right)\Big] dt\nonumber\\ 
    & +\left[\tfrac{\sqrt{\gamma}}{2}(\hat{M} - \ex{\hat{M}})
 -\tfrac{i}{\hbar \sqrt{\gamma}}\hat{F}\right]dW \Big\} \ket{\psi}\,.
\end{align} 
It is possible to formulate this non-deterministic state change in terms of a generator as
\begin{align}
     \ket{\psi} + d\ket{\psi} =   \exp(\hat{G}) \ket{\psi}\,,\label{eq:psi_dpsi_generator}
\end{align}
where 
\begin{align}
\hat{G} :=& \Big[-\tfrac{i}{\hbar}\hat{H}  -\tfrac{\gamma}{4}(\hat{M} - \ex{\hat{M}})^2  + \tfrac{i}{4\hbar} \left[\hat{M},\hat{F}\right] -\tfrac{i}{\hbar} \ex{\hat{M}}\hat{F}\Big] dt\nonumber\\ 
    & \,+\Big[\tfrac{\sqrt{\gamma}}{2}(\hat{M} - \ex{\hat{M}})  -\tfrac{i}{\hbar \sqrt{\gamma}}\hat{F} \Big]dW \,.\label{eq:G}
\end{align}
Up to this point we have not placed any restrictions on the Hamiltonian, $\hat{H}$, the measured observable, $\hat{M}$, or the generator of feedback, $\hat{F}$. Equation~\eqref{eq:psi_dpsi_generator} is therefore valid for any system undergoing continuous measurement of an observable $\hat{M}$ and feedback of the form \eqref{eq:unitary_operator} with hermitian feedback operator $\hat{F}$.


As outlined by Diosi in \cite{diosi1988localized}, we now consider the dynamics of the system in the co-moving frame, i.e.~the reference frame in which the expectation values of both position and momentum are zero. 
For this purpose,  we transform the state $\ket{\psi} + d\ket{\psi} $ at time $t+dt$ into the co-moving frame as follows
\begin{align}
    \ket{\tilde{\psi}} + d\ket{\tilde{\psi}} =  e^{-\frac{i}{\hbar} (\ex{\hat{p}}+d\ex{\hat{p}}) \hat{q}}  
    e^{\frac{i}{\hbar} (\ex{\hat{q}}+d\ex{\hat{q}}) \hat{p}} \left( \ket{\psi} + d\ket{\psi}\right)\,, \label{eq:displacement}
\end{align}  
where $d\ex{\hat{q}} = \mathrm{Tr}(\hat{q}d\rho)$ and $d\ex{\hat{p}} = \mathrm{Tr}(\hat{p}d\rho)$. Transformation \eqref{eq:displacement} resembles feedback however, it is cannot be realised in a laboratory and therefore does not represent standard feedback, rather a purely mathematical operation.  As a consequence of Eq.~\eqref{eq:displacement}, the expectation value of position with respect to the transformed state at time a $t+dt$ vanishes:
\begin{align*}
&(\bra{\tilde{\psi}} + d\bra{\tilde{\psi}})\hat{q}(\ket{\tilde{\psi}} + d\ket{\tilde{\psi}})\nonumber \\ 
&= ( \bra{\psi} + d\bra{\psi})e^{-\tfrac{i}{\hbar} (\ex{\hat{q}}+d\ex{\hat{q}}) \hat{p}} \hat{q} e^{\tfrac{i}{\hbar} (\ex{\hat{q}}+d\ex{\hat{q}}) \hat{p}}( \ket{\psi} + d\ket{\psi})\\
& = ( \bra{\psi} + d\bra{\psi})\left[\hat{q} -(\ex{\hat{q}}+d\ex{\hat{q}}) \right]( \ket{\psi} + d\ket{\psi})   \\
& = 0\,.
\end{align*}
Similarly the expectation value of momentum, $(\bra{\tilde{\psi}} + d\bra{\tilde{\psi}})\hat{p}(\ket{\tilde{\psi}} + d\ket{\tilde{\psi}})$ equals zero. The state $\ket{\tilde{\psi}} + d\ket{\tilde{\psi}}$ therefore represents the state of the system viewed from the co-moving reference frame. 


It is always possible to choose a special coordinate system where $\ex{\hat{q}}=0=\ex{\hat{p}}$ at the given instant $t$ when $\ket{\psi}$ is considered. At this instant $\ket{\psi}$ coincides with $\ket{\tilde{\psi}}$ and it follows that the state change with respect to this coordinate system reads
\begin{align}
   \ket{\tilde{\psi}} + d\ket{\tilde{\psi}}  =&e^{-\frac{i}{\hbar} d\ex{\hat{p}} \hat{q}} e^{\frac{i}{\hbar} d\ex{\hat{q}} \hat{p}}  e^{ \hat{G}} \ket{\tilde{\psi}} \,.\label{eq:psi_dpsi}
\end{align}

Note, that we have not placed any restrictions on the Hamiltonian, $\hat{H}$, measured observable, $\hat{M}$, or the generator of feedback, $\hat{F}$. 
Therefore, by replacing $q$ and $p$  in Eq.~(\ref{eq:psi_dpsi}) with any pair of conjugate variables, we can shift the description of the system into a reference frame with respect to which the expectation values of these conjugate variables vanish simultaneously. Using the measured observable $M$ and its conjugate momentum as the conjugate variables, greatly simplifies the calculation of the stationary states as we shall see in Section~\ref{sec:stat_har}.  This simplification is the essence of Diosi's method, and the form invariance of  Eq.~(\ref{eq:psi_dpsi}) under canonical transformations makes it applicable to compute the stationary states for a large class of continuous measurements with feedback.

\section{Stationary state of a particle in a harmonic potential}\label{sec:stat_har}

In order to illustrate the power of this generalised approach we consider the specific case that the measured observable is an arbitrary linear combination of position and momentum, 
\begin{align}
\hat{M }= \alpha \hat{q} + \beta \hat{p} \,,
\end{align}
where $\alpha$ and $\beta$ are real constants of units $1$ and $\si{\per\kilo\gram\second}$, respectively. Furthermore, let the generator of feedback be 
\begin{align}\label{eq:F_chidelta}
\hat{F} = \chi \hat{q} + \delta \hat{p}
\end{align}
where $\chi$ and $\delta$ are real gain factors of units $\si{\kilo\gram\second^{-2}}$ and $\si{\per\second}$, respectively. In order to simplify the calculations that follow, we introduce the following operators 
\begin{align}
\hat{Q}&:= \hat{M} =  \alpha \hat{q} +\beta \hat{p} \label{eq:Q}\\
\hat{P}&:= - \beta' \hat{q} +\alpha' \hat{p} \label{eq:P_alpha_beta}
\end{align}
that satisfy the canonical commutation relation, $[\hat{Q},\hat{P}]=i\hbar $, where for simplicity we assume that the factor ${\alpha\alpha'+ \beta\beta'} =1$. This assumption also implies that $Q$ and $P$ are canonically conjugate. In addition, we can write
\begin{align}\label{eq:F_uv}
\hat{F} = u \hat{Q} + v \hat{P}\,,
\end{align}
where $u$ and $v$ are real factors of units $\si{\kilo\gram\second^{-2}}$ and $\si{\per\second}$, respectively. 
The transition to the co-moving frame can also be achieved by replacing variables $q$ ($p$) in transformation (\ref{eq:displacement}) by the conjugate pair $Q$ and $P$ since the expectation values of position and momentum vanish if, and only if, those of the generalised variables $Q$ and $P$ vanish.

As an example, let us consider a particle in a harmonic potential with Hamiltonian
\begin{align}
\hat{H} = \frac{1}{2m} \hat{p}^2 + \frac{m\omega^2}{2}\hat{q}^2\,.
\end{align}
If $\alpha'=\alpha$ and $\beta' = m^2\omega^2\beta$, the Hamiltonian of a harmonic oscillator is form invariant under linear transformations of this type, so that
\begin{align}
\hat{H} = \frac{1}{2m} \hat{P}^2 + \frac{m\omega^2}{2}\hat{Q}^2\,.\label{eq:H_QP}
\end{align}

We now show that the transformation~\eqref{eq:psi_dpsi} for a particle in a harmonic potential leads to a unique solution in the stationary regime.  Applying the Baker-Campbell-Hausdorff formula and the Ito rule \eqref{eq:Ito_rules} to Eq.~\eqref{eq:psi_dpsi} results in a state change in the co-moving frame given by 
\begin{align}
   &\ket{\tilde{\psi}}  + d\ket{\tilde{\psi}}= \nonumber \\& \exp\Big\{  \left[ - \tfrac{i}{\hbar}\hat{{H}} - \tfrac{\gamma}{4}\left(\hat{Q}^2 - \ex{\hat{Q}^2}\right)\right] dt \nonumber \\ & \hspace{18pt}-\tfrac{i}{\hbar}\sqrt{\tfrac{\gamma}{2}} \Big[ \left(i\hbar + \ex{\{\hat{Q},\hat{P}\}} \right)\hat{Q} - 2\ex{\hat{Q}^2} \hat{P}\Big] dW \Big\} \ket{\tilde{\psi}} ,\label{eq:QLE}
\end{align} 
where we ignored irrelevant phase factors and assumed Hamiltonian~\eqref{eq:H_QP}. The full derivation and the general state change for arbitrary Hamiltonian are provided in Appendix~\ref{app:co-moving}.

We note that Eq.~\eqref{eq:QLE} no longer depends on feedback. This is to be expected as feedback of form \eqref{eq:unitary_operator} corresponds to a shift of the Wigner function \cite{olivares2012quantum} in phase space, and consequently its shape remains unchanged. Therefore, from the point of view of the co-moving frame, feedback has no impact on the state of the quantum particle. We note that if the generator of feedback $\hat{F}$ contains terms of higher orders of position and momentum, for example angular momentum, then this is no longer true. However, this method still leads to a simplified state change in the co-moving frame since all first-order terms in Q and P are eliminated. 

The stationary solutions are of the form $\ket{\tilde{\psi}_\infty} \exp(-iEt/\hbar)$, where $\ket{\tilde{\psi}_\infty}$ is the time-independent part and $E$ is a real number with units of energy. Substitution of this ansatz into Eq.~\eqref{eq:QLE} followed by a comparison of coefficients of the independent increments $dt$ and $dW$ yields 
\begin{align}
\left[ \hat{{H}}  -i \tfrac{\hbar \gamma}{4}\left(\hat{Q}^2 - \ex{\hat{Q}^2}_\infty\right)\right]  \ket{\tilde{\psi}_\infty} = E \ket{\tilde{\psi}_\infty} \,,\label{eq:terms_dt}
\end{align}
such that $E= \braket{\tilde{\psi}_{\infty} |\hat{H}|\tilde{\psi}_\infty} \equiv \ex{\hat{H}}_{\infty}$, and
\begin{align}
\Big[ \left(i\hbar + \ex{\{\hat{Q},\hat{P}\}}_\infty \right)\hat{Q} - 2\ex{\hat{Q}^2}_\infty \hat{P}\Big] \ket{\tilde{\psi}_\infty}  = 0\,. \label{eq:terms_dW}
\end{align}
If we introduce the wave function in position representation $\tilde{\psi}(q)$, Eq.~\eqref{eq:terms_dW} becomes a first-order linear ordinary differential equation of the form $aq \tilde{\psi}(q) = {d\tilde{\psi}(q)}/{dq}$, where $a$ is a complex constant, namely
\begin{align}
\big(i\hbar\alpha + \ex{\{\hat{Q},\hat{p}\}}_\infty& \big){q}\tilde{\psi}_\infty(q) \nonumber \\& = i\hbar \big(i\hbar\beta - \ex{\{\hat{Q},\hat{q}\}}_\infty \big)\frac{d\tilde{\psi}_\infty(q)}{dq}   \,. \label{eq:terms_dW_qp}
\end{align}
This differential equation is known to have a Gaussian solution and it follows that the stationary solution in position representation is  given by
\begin{align}
\tilde{\psi}_\infty(q) = & \left(\frac{\operatorname{Re}\{\sigma^2\}}{2\pi|\sigma^2|^2}\right)^{\frac{1}{4}}
\exp\left( -  \frac{q^2}{4\sigma^2} \right)\,,
\end{align}
where
\begin{align}
\sigma^2 &:= \frac{i\hbar}{2} \frac{ \ex{\{\hat{q},\hat{Q}\}}_\infty-i\hbar\beta}{ \ex{\{\hat{p},\hat{Q}\}}_\infty +i\hbar \alpha}\,,
\end{align}
and similarly in momentum space
\begin{align}
\tilde{\psi}_\infty(p) = &\left(\frac{2\operatorname{Re}\{\sigma^2\}}{\pi \hbar^2}\right)^{\tfrac{1}{4}}\exp\left(- \frac{\sigma^2 p^2}{\hbar^2} \right)\,.
\end{align}
The eigenvalue equation~\eqref{eq:terms_dW}, and equivalently Eq.~\eqref{eq:terms_dW_qp},  holds for arbitrary Hamiltonian $\hat{H}$, see Eq.~\eqref{ap:state_change_genH}. Therefore, one could ask the question whether arbitrary Hamiltonians allow the energy eigenvalue equation
and eigenvalue equation \eqref{eq:terms_dW}, that follow from Eq.~\eqref{ap:state_change_genH}, to be simultaneously satisfied. For the free particle it is possible to use the results of \cite{diosi1988localized} to show that both eigenvalue equations hold. We briefly argue that this is also the case for a general Hamiltonian however, the analysis is be beyond the scope of this paper. 

For the harmonic oscillator we can show explicitly that these Gaussian stationary wave functions also satisfy the energy eigenvalue equation \eqref{eq:terms_dt} for all measurements strengths $\gamma$, see Appendix~\ref{ap:eigen}. Proving this requires us to express the stationary value of the variance of the measured observable in the co-moving frame, $\ex{\hat{Q}^2}_\infty$, as a function of the expectation value of the anti-commutator  $\ex{\{\hat{Q}, \hat{P}\}}_\infty$ and the variance $\ex{\hat{P}^2}_\infty$. The calculation of these terms is the subject of the next section. It turns out that under certain circumstances the quantities $\ex{\hat{Q}^2}$, $\ex{\hat{P}^2}$ and $\ex{\{\hat{Q},\hat{P}\}}$ obey a set of coupled first-order differential equations and thus evolve deterministically. Further, in Appendix~\ref{ap:eigen} it is shown that for a particle in a harmonic potential the energy eigenvalue in Eq.~\eqref{eq:terms_dt} is given by
\begin{align}
E = &\frac{\hbar^2}{4m \ex{\hat{Q}^2}_\infty}
\,.\label{eq:energy_stationary}
\end{align}
It also turns out that in the weak measurement limit, $\gamma\rightarrow 0$, the energy with respect to the co-moving frame is minimal, i.e.,  $\ex{\hat{Q}^2}_\infty\rightarrow \frac{\hbar}{2m\omega}$. In addition, the product $\ex{\hat{Q}^2}_\infty \ex{\hat{P}^2}_\infty \rightarrow \frac{\hbar^2}{4}$ so that $E \rightarrow \ex{\hat{P}^2}_\infty/m$, which is twice the expectation value of the kinetic energy in agreement with the Virial theorem.

\subsection{Calculation of stationary widths for a particle in a harmonic potential}
 From the expressions of second-order moments of position and momentum a system of differential equations can be derived that allows to calculate the width of the asymptotic wave function, 
\begin{align}
d\ex{\Delta \hat{q}^2}  = &  d\ex{\hat{q}^2} - 2\ex{\hat{q}}d\ex{\hat{q}} - (d\ex{\hat{q}})^2\label{eq:dq2}\\
d\ex{\Delta \hat{p}^2} = & d\ex{\hat{p}^2} - 2\ex{\hat{p}}d\ex{\hat{p}} - (d\ex{\hat{p}})^2\label{eq:dpq}\\
d\ex{\{\Delta \hat{q}, \Delta \hat{p}\}} = & d\ex{\{\hat{q},\hat{p}\}} - 2\ex{\hat{q}}d\ex{\hat{p}} - 2\ex{\hat{p}}d\ex{\hat{q}} \nonumber \\ & - 2d\ex{\hat{q}}d\ex{\hat{p}}\label{eq:dp2}\, .
\end{align}
Here we have introduced the notation $\Delta \hat{o} = \hat{o} - \ex{\hat{o}}$, such that the variance of $\hat{o}$ is given by $\ex{\Delta \hat{o}^2} = \ex{\hat{o}}^2-\ex{\hat{o}^2}$. Note that the last terms in Eqs.~\eqref{eq:dq2}-\eqref{eq:dp2} take into account the stochastic contribution $dW$ in the increment of the expectation value, see equation below, the square of which leads to a contribution of order $dt$. The increment $d\ex{\hat{o}}$ of an arbitrary operator $\hat{o}$ can be calculated using the master equation~\eqref{eq:stochastic_master},
\begin{align}\label{eq:da}
d\ex{\hat{o}} =& \mathrm{Tr}(\hat{o}d\rho) \nonumber\\
 =& \Big[-\tfrac{i}{\hbar}\ex{[\hat{o},\hat{H}]} -\tfrac{\gamma}{8}\ex{[\hat{Q},[\hat{Q},\hat{o}]]}  \nonumber \\ & \quad -\tfrac{1}{2\hbar^2\gamma}\ex{[\hat{F},[\hat{F},\hat{o}]]} -\tfrac{i}{2\hbar}\ex{\{\hat{Q},[\hat{o}, \hat{F}]\}}\Big]dt\nonumber \\ &  + \Big[ \tfrac{\sqrt{\gamma}}{2} \ex{\{ \hat{o} , \Delta \hat{Q}\}} -\tfrac{i}{\hbar \sqrt{\gamma}} \ex{[\hat{o},\hat{F}]}\Big] dW\,.
\end{align}

Equations \eqref{eq:dq2}-\eqref{eq:dp2} are form invariant for any pair of conjugate variables. We can thus compute these total differentials for $\hat{Q} $ and $\hat{P}$. 
\begin{align}
d\ex{ (\Delta \hat{Q})^2}  = & -\gamma\ex{ (\Delta \hat{Q})^2}^2dt 
\nonumber \\& -\frac{i}{\hbar} \ex{\{\Delta \hat{Q}, \Delta[\hat{Q},\hat{H}]\}} dt \nonumber \\&+{\sqrt{\gamma}} \ex{(\Delta \hat{Q})^3} dW \label{eq:td_1}\\
d\ex{ (\Delta \hat{P})^2} =  &\frac{\hbar^2\gamma}{4}dt  -\frac{\gamma}{4}\ex{\{\Delta \hat{P}, \Delta \hat{Q} \}}^2dt \nonumber \\&-\frac{i}{\hbar} \ex{\{\Delta \hat{P}, \Delta[\hat{P},\hat{H}]\}} dt\nonumber \\& +\tfrac{\sqrt{\gamma}}{2} \ex{\{ (\Delta \hat{P})^2, \Delta \hat{Q}\}} dW \label{eq:td_2}\\
d\langle\{\Delta \hat{Q}, \Delta \hat{P} \}\rangle = &  -\frac{i}{\hbar}  \ex{\{\Delta \hat{Q}, \Delta[\hat{P},\hat{H}]\}}dt \nonumber \\ &  -\frac{i}{\hbar} \ex{\{\Delta \hat{P}, \Delta[\hat{Q},\hat{H}]\}} dt\nonumber \\& - \gamma \ex{\{\Delta \hat{P}, \Delta \hat{Q} \}}\ex{ (\Delta \hat{Q})^2}dt \nonumber \\& +\tfrac{\sqrt{\gamma}}{2} \ex{\{ \{\Delta \hat{Q}, \Delta \hat{P} \}, \Delta \hat{Q} \}} dW \,.\label{eq:td_3}
\end{align}
This set of coupled equations holds for all times $t$ and for an arbitrary Hamiltonian, $H$. 

For a particle in a harmonic potential and for large times $t >> \frac{m\omega}{\hbar\gamma}$, we can use the fact that the stationary states have a Gaussian form, see Section~\ref{sec:stat_har}. This also applies  for a free particle and it can also be assumed in good approximation whenever the particle is localised due to the measurement on length scales on which the potential does not vary much \cite {halliwell1995quantum}. In this case higher order moments are zero, in particular
 \begin{align}
 \ex{\Delta \hat{Q}^3 }_\infty& = \ex{\{(\Delta \hat{P})^2 , \Delta \hat{Q} \}}_\infty \nonumber \\&= \ex{\{\{\Delta \hat{Q},\Delta \hat{P}\}, \Delta \hat{Q} \}}_\infty =0\,.\label{eq:higher_order_var}
 \end{align} 

\begin{figure}
     \centering
     \includegraphics[width=\columnwidth]{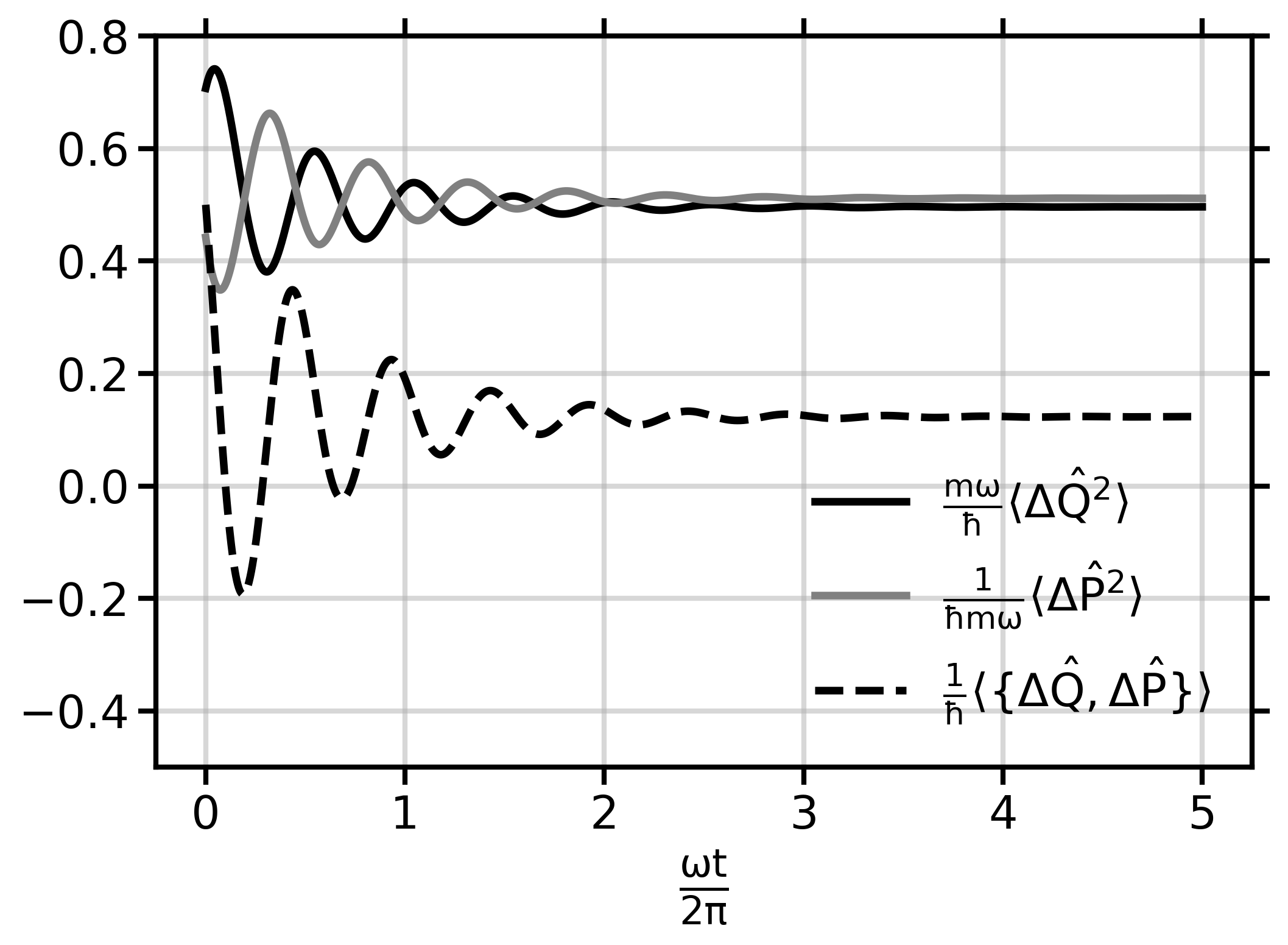}
     \caption{Time dependant solutions to the coupled differential equations~\eqref{eq:dx}, \eqref{eq:dy} and \eqref{eq:dz} for $\kappa=1/4$. The initial conditions were ${\frac{m\omega}{\hbar}}\langle \hat{Q}^2 \rangle = 1/\sqrt{2}$ and $\frac{1}{\hbar}\langle \{\Delta\hat{Q},\Delta\hat{P}\}\rangle = 1/2$, with ${\frac{1}{\hbar m\omega}}\langle \hat{P}^2 \rangle = 5/(8\sqrt{2})$ according to Eq.~\eqref{eq:xyz}.}
     \label{fig:21}
\end{figure}

Assuming that the Hamiltonian $\hat{H}$ is that of the harmonic oscillator \eqref{eq:H_QP}, we note that $[\hat{Q}, \hat{H}] =\tfrac{i\hbar}{m} \hat{P}$  and $[\hat{P}, \hat{H}] =   -i\hbar m\omega^2\hat{Q}$ so that Eqs.~\eqref{eq:td_1}-\eqref{eq:td_3} are  reduced to a system of linear  first-order coupled differential equations which can be written as
\begin{align}
\dot{x}& = -2\kappa x^2 + z \label{eq:dx}\\
\dot{y}& = \frac{\kappa}{2}\left(1-z^2\right)-z\label{eq:dy}\\
\dot{z}& = 2\left(y-x\right)-2\kappa x z\label{eq:dz}\,,
\end{align}
where 
\begin{align}
x:= & \frac{m\omega}{\hbar}\ex{ (\Delta \hat{Q})^2} \label{eq:x}\\
y:= & \frac{1}{\hbar m \omega}\ex{(\Delta  \hat{P})^2}\label{eq:y}\\
z:= & \frac{1}{\hbar}\ex{\{\Delta \hat{Q}, \Delta \hat{P} \}}\label{eq:z}
\end{align}
are the unitless versions of the variances of interest and
\begin{align}
\kappa & := \frac{\hbar \gamma }{2m\omega^2 } \label{eq:kappa_dum}
\end{align} 
is the measurement strength relative to the characteristic dynamics of the particle. The steps taken to arrive at this result are shown in Appendix~\ref{Appendix:dimensionless}. The coupled differential equations \eqref{eq:dx}-\eqref{eq:dz} can be solved numerically, with an example of the numerically obtained solutions for specific initial conditions shown in Figure~\ref{fig:21}. In addition, we note that Eqs.~\eqref{eq:dx}-\eqref{eq:dz} imply 
\begin{align}\label{eq:xyz}
 \ex{ (\Delta \hat{Q})^2}\ex{ (\Delta \hat{P})^2} = \frac{\hbar^2}{4} + \frac{1}{4}
 \ex{\{\Delta \hat{Q}, \Delta \hat{P} \}}^2 \,,
\end{align}
so that $\ex{\{\Delta \hat{Q}, \Delta \hat{P} \}}$ increases the minimal uncertainty~\cite{sakurai2014modern}.

In the stationary regime the time derivatives $d\ex{ (\Delta \hat{Q})^2}/dt$, $d\ex{ (\Delta \hat{P})^2}/dt$ and $d\ex{ \{\Delta \hat{Q},\hat{P}\}}/dt$ (equivalently $\dot{x}$, $\dot{y}$ and $\dot{z}$) are zero, so that Eqs.~\eqref{eq:td_1}-\eqref{eq:td_3} (Eqs.~\eqref{eq:dx}-\eqref{eq:dz}) yield
\begin{align}
\frac{1}{\hbar}\ex{\{\Delta \hat{Q}, \Delta \hat{P} \}}_\infty 
& = \frac{1}{\kappa}\left(- 1 + \sqrt{1+\kappa^2}\right)\label{eq:anti_com_inf_kappa}\\
\frac{m\omega}{\hbar}\ex{ (\Delta \hat{Q})^2}_\infty 
&=\frac{1}{\sqrt{2}\kappa} \sqrt{- 1 + \sqrt{1+\kappa^2}}\label{eq:Q2_inf_kappa}\\
\frac{1}{\hbar m \omega}\ex{(\Delta  \hat{P})^2}_\infty    
&= \frac{\sqrt{1+\kappa^2}}{\sqrt{2}\kappa} \sqrt{- 1 + \sqrt{1+\kappa^2}} \label{eq:P2_inf_kappa}\,.
\end{align}
The above solutions are valid for all values of $\gamma$, $\alpha$, $\beta$, $m$ and $\omega$ provided that the assumption $\alpha^2+m^2\omega^2\beta^2 = 1$ holds. We note that the asymptotic values depend on $m\omega$ and the dimensionless relative measurement strength $\kappa$, and not on $m$, $\omega$ and $\gamma$ independently. 

\begin{figure}
         \includegraphics[width=\columnwidth]{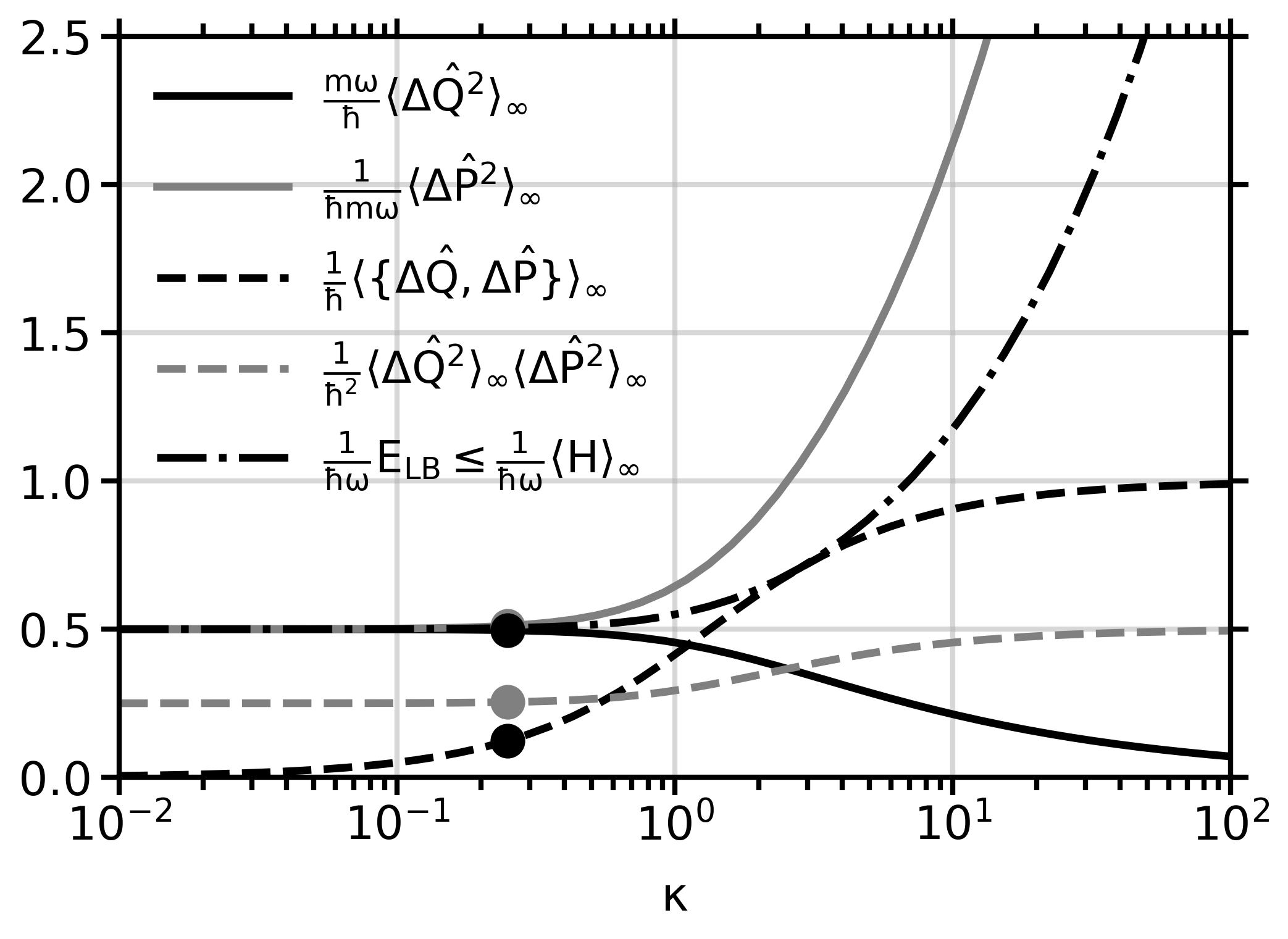}
         \caption{Plot showing the stationary value of the second-order variances as a function of $\kappa$ according to Eqs.~\eqref{eq:anti_com_inf_kappa}-\eqref{eq:P2_inf_kappa}. The dashed grey line represents the asymptotic uncertainty as a function of $\kappa$. A lower bound for the asymptotic energy $E_{\text{LB}}$ is also shown and is determined by the expectation value of the Hamiltonian with respect to the stationary state in the co-moving frame~\eqref{eq:energy_stationary}. Dots indicate the stationary values associated with $\kappa = 0.25$ for comparison with Figure~\ref{fig:21}.}
\label{fig:stationary_quantities}
\end{figure}

The relative measurement strength $\kappa$ is the product of the measurement strength $\gamma$ and the two quantities that characterise the dimensions of the system. These are the time scale on which the system oscillates $1/\omega$, and the square length scale given by the variance $\frac{\hbar}{2m\omega}$ of the position of the ground state of the harmonic oscillator. If $\kappa >> 1$, measurement is the dominant feature and the particle follows a random walk. If, on the other hand, $\kappa << 1$ the  measurement is weak compared to the unitary dynamics of the system and the harmonic motion of the particle is dominant.  The asymptotic second-order moments \eqref{eq:anti_com_inf_kappa}-\eqref{eq:P2_inf_kappa} are plotted as functions of the relative measurement strength $\kappa$ in Figure~\ref{fig:stationary_quantities}, which shows the transition that occurs in the regime $1 < \kappa < 10$ from a state of minimal uncertainty to a state of maximal uncertainty that the system can assume. 

In the weak measurement limit $\gamma \rightarrow 0$, i.e.~$\kappa \rightarrow 0$, the variance of the measured observable approaches the variance of the ground state,
\begin{align}
\ex{ (\Delta \hat{Q})^2}_\infty \rightarrow \frac{\hbar}{2m\omega}\,.
\end{align}
Therefore, according to Eq.~\eqref{eq:energy_stationary}, the energy in the co-moving frame converges to the ground state energy  $\hbar \omega/2$, as alluded to in the discussion of Eq.~\eqref{eq:energy_stationary}. Values of $\kappa < 1$ are experimentally feasible for systems such as trapped ions and optical oscillators. 

\section{Feedback for cooling and confinement}\label{section:Analysing_feedback}
The results of the previous section are dictated by measurement only, since the feedback vanishes from the equation of motion in the co-moving frame. Here we consider the effect of  feedback on the motion of the quantum particle viewed from the lab frame and how it can be used to cool and confine the particle. 

At each moment in time, feedback changes the position and momentum of the particle by an amount proportional to the measurement signal and according to coefficients $\chi$ and $\delta$ in Eq.~\eqref{eq:F_chidelta}. By tuning these coefficients, we can alter the motion of the particle in such a way as to mimic the action of a quadratic potential. In general, this allows for the Hamiltonian of the system to be altered by a term quadratic in the measured observable.  

Feedback can also be used for the purpose of cooling and confinement. Intuitively, information gained from measurement can be used to gradually reduce the expectation value of the measured observable $Q$ by shifting the system towards a smaller value of this observable. This is achieved by continuously applying feedback generated by the conjugate momentum $P$ with a negative scaling. At the same time, it is possible to choose this scaling such that the random changes in $Q$ introduced by the measurement diminish over time. The same can be achieved for the conjugate momentum $P$. This is realised by introducing friction generated by $Q$ which simultaneously compensates for random changes of the conjugate momentum $P$ generated by the measurement of $Q$. 
The combined feedback thus leads to damping of both the measured observable $Q$ and its conjugate momentum $P$. To determine the values of the scaling coefficients $u$ and $v$ in Eq.~\eqref{eq:F_uv} that lead to optimal cooling and confinement, it is necessary to consider the equations that govern the motion of the particle.

The stochastic master equation \eqref{eq:stochastic_master} in terms of the operators $\hat{Q}$ and $\hat{P}$ reads
\begin{align}\label{eq:gen_fb}
 d\rho = &-\tfrac{i}{\hbar}\left[\hat{H} +\tfrac{u}{2}\hat{Q}^2,\rho\right]dt
  \nonumber \\
& - \tfrac{\gamma}{8}\left(1+\left(\tfrac{2u}{\hbar\gamma}\right)^2\right)\left[\hat{Q},\left[\hat{Q},\rho\right]\right] dt \nonumber \\
& -\tfrac{v^2}{2\hbar^2\gamma}\left[\hat{P},\left[\hat{P},\rho\right]\right] dt - \tfrac{iv}{2\hbar}\left[\hat{P}, \left\{\hat{Q},\rho\right\}\right]  dt \nonumber \\ &-\tfrac{uv}{2\hbar^2\gamma}\left(\left[\hat{Q},\left[\hat{P},\rho\right]\right]+\left[\hat{P},\left[\hat{Q},\rho\right]\right]\right) dt \nonumber \\
&  + \tfrac{\sqrt{\gamma}}{2}\left\{\hat{Q}- \langle \hat{Q}\rangle,\rho\right\} dW - \tfrac{i}{\hbar\sqrt{\gamma}}\left[u\hat{Q}+v\hat{P},\rho\right]dW\,,
\end{align}
where we inserted $\hat{F}=u\hat{Q}+v\hat{P}$, and used the identity $[\hat{Q},\{\hat{Q},\rho\}] = [\hat{Q}^2,\rho]$ to obtain the first term in the this equation. Eq.~\eqref{eq:gen_fb} shows that by choosing $u\neq 0$, the Hamiltonian of the system can be altered by a quadratic term in the measured observable $Q$. This is discussed in Section~\ref{sec:H_mani}.

Furthermore, according to Eq.~\eqref{eq:da} the increments in the measured observable and its conjugate momentum are given by
\begin{align}\label{eq:dQ_uv}
d\ex{\hat{Q}} = & -\tfrac{i}{\hbar}\ex{[\hat{Q},\hat{H}]}dt + v\ex{\hat{Q}}dt \nonumber\\ &+ \sqrt{\tfrac{2\hbar\kappa}{m}}\left(\tfrac{m\omega}{\hbar}\ex{(\Delta\hat{Q})^2} +\tfrac{1}{2\kappa}\tfrac{v}{\omega}\right)dW
\end{align}
and
\begin{align}\label{eq:dP_uv}
d\ex{\hat{P}} = & -\tfrac{i}{\hbar}\ex{[\hat{P},\hat{H}]} dt -u\ex{Q}dt\nonumber\\ & + \sqrt{\tfrac{\hbar m\omega^2 \kappa}{2}}\left(\tfrac{1}{\hbar}\ex{\{\Delta\hat{Q},\Delta\hat{P}\}} -\tfrac{1}{\kappa}\tfrac{u}{m\omega^2}\right)dW \,,
\end{align}
respectively. These equations indicate how to simultaneously eliminate the stochastic terms. As discussed in Section~\ref{sec:cooling}, the corresponding choice of feedback turns out to be sufficient for cooling and confinement of a particle in a harmonic potential. Furthermore, it is possible to drive the particle into the ground state by combining this choice of feedback with weak measurement.


\subsection{Manipulation of the Hamiltonian}\label{sec:H_mani}

First we consider the implications on the system dynamics if we choose feedback in order to alter the Hamiltonian of the system. If the generator of feedback is a linear combination of position and momentum, see Eq.~\eqref{eq:F_chidelta} and Eq.~\eqref{eq:F_uv}, then feedback proportional to the measured observable $Q$ allows us to simulate an additional quadratic potential in $Q$. This is implied by the form of the first term on the right-hand side of Eq.~\eqref{eq:gen_fb}.

In the case of a harmonic potential \eqref{eq:H_QP}, we consider how feedback might be used to compensate for a portion of the Hamiltonian. Let us assume that $\hat{F} = -m\omega^2 \hat{Q}$, i.e. $u=-m\omega^2$ and $v=0$. The master equation \eqref{eq:gen_fb} becomes
\begin{align}\label{eq:drho_P_uq}
 d\rho = &-\tfrac{i}{\hbar}\left[\hat{H} - \tfrac{m\omega^2}{2}\hat{Q}^2,\rho\right]dt
  \nonumber \\
& - \tfrac{\gamma}{8}\left(1+\left(\tfrac{2m\omega^2}{\hbar\gamma}\right)^2\right)\left[\hat{Q},\left[\hat{Q},\rho\right]\right] dt \nonumber \\
&  + \tfrac{\sqrt{\gamma}}{2}\left\{\hat{Q}- \langle \hat{Q}\rangle,\rho\right\} dW + \tfrac{im\omega^2}{\hbar\sqrt{\gamma}}\left[\hat{Q},\rho\right]dW\,.
\end{align}
Here feedback cancels a portion of the Hamiltonian so that the effective Hamiltonian $\hat{H}_{\text{eff}}$ is given by
\begin{align}
\hat{H}_{\text{eff}} &:= \hat{H} - \tfrac{m\omega^2}{2}\hat{Q}^2 \nonumber \\ &= \tfrac{1}{2m}\hat{P}^2 
\end{align}
The master equation~\eqref{eq:drho_P_uq} resembles that of a free particle under continuous position measurement, with an amplified decoherence term, as well as a random Hamiltonian (cp.~last term in the equation). 

To take a particular example, let us continuously measure position, with $\hat{Q}=\hat{q}$ and $\hat{P}=\hat{p}$ (i.e. $\alpha = 1$ and $\beta =0$), so that $\hat{H}_{\text{eff}}$ is the Hamiltonian of a free particle. According to Eq.~\eqref{eq:dQ_uv} and in the non-selective regime of measurement, the expectation value of the position $q$ grows linearly proportional to the expectation value of the conjugate momentum,
while the expectation value of the  momentum ${p}$ remains constant, according to Eq.~\eqref{eq:dP_uv}. A method to continuously infer the position of the ion through spontaneous light scattering into mirror modes is provided in \cite{steixner2005quantum}. 

Alternatively, let us continuously monitor momentum, with $\hat{Q}=(m\omega)^{-1}\hat{p}$ and $\hat{P}=-m\omega\hat{q}$, i.e. $\alpha=0$ and $\beta=(m\omega)^{-1}$ in order to satisfy the condition $\alpha^2+m^2\omega^2\beta^2=1$. The measurement of the momentum of the ion can be realised via the measurement of the phase changes of a probe beam using homodyning\cite{rabl2005quantum}. For this choice of feedback the effective Hamiltonian $\hat{H}_{\text{eff}}$ is given by $\frac{m\omega^2}{2}\hat{q}^2$. For this choice of feedback and in the non-selective regime of measurement, the momentum $p$ grows linearly proportional to the position $q$, while the position remains constant, according to Eq.~\eqref{eq:dQ_uv} and Eq.~\eqref{eq:dP_uv} respectively.  Feedback continuously shifts the quantum particle back to its original position, while it simultaneously gains momentum as a result of the force applied by the harmonic potential.

\subsection{Cooling and confinement of a particle in a harmonic potential}\label{section:temp}
In order to investigate the thermalising effect of the feedback, we first compare the master equation \eqref{eq:gen_fb} with that of an optical harmonic oscillator coupled to a heat bath. We consider the prototypical example of an optical oscillator in the rotating wave approximation and in the non-selective regime of measurement, allowing us to compare our results with the typical equations obtained for such a system. 

In the second part of this section we considering the equations governing the increment of the energy, as well as the expectation values of the measured observable and the conjugate momentum, outside of the limits of the aforementioned approximations. This allows us to derive a generalised approach to cooling that facilitates the simultaneous  reduction of the expectation values of both position and momentum at an exponential rate. 

\subsubsection{Optical oscillator coupled to a heat bath}\label{section:temp_optical}
The position and momentum operators, $\hat{q}$ and $\hat{p}$, can be expressed in terms of creation and annihilation operators as,
\begin{equation}\label{eq:newqp}
  \hat{q} = \sqrt{\frac{\hbar}{2m\omega}} (\hat{a}+\hat{a}^\dagger) \;\;\; \mbox{and} \;\;\;\hat{ p} = i\sqrt{\frac{\hbar m\omega}{2}} (\hat{a}^\dagger-\hat{a}).
\end{equation}
In Appendix~\ref{AppendixC} the master equation \eqref{eq:stochastic_master} written in terms of creation and annihilation operators is given in Eq.~\eqref{eq:gen_creation}.  We consider this result in the rotating wave approximation and in the non-selective regime of measurement, such that the state of an optical harmonic oscillator obeys
 \begin{align} \label{tempeq}
   d\rho' =& -\tfrac{i}{\hbar}\left[\hat{H},\rho\right]dt  - \tfrac{i u}{4 m\omega}[\left\{\hat{a},\hat{a}^\dagger\right\},\rho]dt \nonumber \\
  &+ \left(c-v \right) (\hat{a}\rho\hat{a}^\dagger-\tfrac{1}{2}\{\hat{a}^\dagger\hat{a},\rho\})dt \nonumber \\
  &+ c (\hat{a}^\dagger\rho\hat{a}-\tfrac{1}{2}\{\hat{a}\hat{a}^\dagger,\rho\})dt\,,
\end{align}
where the real constant $c$ is given by
\begin{align}
c = \tfrac{\kappa \omega}{4} +\tfrac{1}{4\kappa \omega} (\tfrac{u^2}{m^2\omega^2}+ v^2) + \tfrac{v}{2} \,.
\end{align}
For a particular class of feedback, namely $\hat{F} \propto \hat{P}$ with $u=0$,  Eq.~\eqref{tempeq} has the same form as the equation governing a harmonic oscillator coupled to a heat bath of harmonic oscillators at a certain temperature\cite{WisemanBook}, given by
\begin{align} \label{lindblad}
d\rho = & - \tfrac{i}{\hbar}[H,\rho] dt \nonumber \\ & + \gamma' (N+1) (\hat{a}\rho\hat{a}^\dagger-\tfrac{1}{2}\{\hat{a}^\dagger\hat{a},\rho\}) dt \nonumber \\
&+ \gamma' \ N \ (\hat{a}^\dagger\rho\hat{a}-\tfrac{1}{2}\{\hat{a}\hat{a}^\dagger,\rho\})dt\,,
\end{align}
where $\gamma'$ is the decay rate and $N = 1/(\exp(\tfrac{\hbar\omega}{k_BT})-1)$ is the mean excitation of the heat bath of harmonic oscillators, with $T$ being the temperature of the bath. The expression for $N$ can be reformulated as
\begin{equation}
T = \frac{\hbar\omega}{k_B\ln((N+1)/N)}\,.\label{temperature}
\end{equation}

A system in thermal contact with a heat bath as described by Eq.~\eqref{lindblad} will eventually be thermalised at temperature $T$.  Because Eq.~\eqref{lindblad} also represents the dynamics of our controlled harmonic oscillator,  appropriate measurement and feedback induce a thermalising effect similar to the coupling to an external heat bath.   By comparing Eqs.~\eqref{tempeq} and \eqref{lindblad}, the conditions for thermalisation in terms of the measurement strength and feedback can be determined. 

From the difference between the coefficients of the pumping and the dissipation terms, we obtain the rate of thermalisation, i.e.~$\gamma' = -v$. If  $\gamma'$ is greater than zero, then the system loses more energy via dissipation than it gains by pumping, which is equivalent to cooling. We can therefore conclude that if the proportionality constant $v$ is negative then energy is reduced, the system is cooled. 

That the generator of feedback should be chosen proportional to the conjugate momentum $P$ with a negative scaling factor is intuitively correct. Doing so correspond to performing a shift of the measured observable towards a smaller value, based on information obtained from measuring the system, as discussed earlier. In Section~\ref{sec:cooling}, we shall see  that in the selective regime of measurement the situation is more subtle, none-the-less the results that follow will be useful for comparison.

The effective temperature of the system can be calculated using the ratio of the coefficients of the pumping and the dissipation terms in Eq.~\eqref{tempeq}. By substituting this ratio into the Eq.~\eqref{temperature}, the effective temperature of the system can be written as
\begin{align}\label{zet}
T & = \frac{\hbar\omega}{2k_B\ln\left(\frac{c-v}{c}\right)} \nonumber \\
&=\frac{\hbar\omega}{2k_B\ln\left(\frac{v-\kappa\omega}{v+\kappa\omega}\right)}\,.
\end{align}
Minimal temperature is therefore achieved if $v = -\kappa\omega$, i.e.~$\hat{F} = -\kappa\omega\hat{P} $, with decay rate $\gamma' = \kappa\omega = \tfrac{\hbar\gamma}{2m\omega}$. As $\gamma/\gamma' \rightarrow 2m\omega/\hbar$, temperature $T \rightarrow 0\,K$. Assuming the control over $\gamma/\gamma'$ to a fourth-order degree i.e, $\gamma/\gamma' = [\tfrac{2m\omega}{\hbar}(1-10^{-4}), \tfrac{2m\omega}{\hbar}(1+10^{-4})]$, for a mechanical system with frequency, $\omega$ of the order of magnitude $10^6$, the temperature of the system is cooled down to the order of $10^{-7}K$. Whereas for an optical system of frequency of the order $10^{12}$, the system is cooled down to a temperature of the order $10^{-1}K$.
 
The dynamics of the mean excitation of the system can be calculated using $\tfrac{d\langle n\rangle}{dt} = -\gamma'd\langle n\rangle+\gamma' N$~\cite{WisemanBook}. This would give 
\begin{equation}
\langle n\rangle = N+(n_i-N)e^{-\gamma't},
\end{equation}
where $n_i$ is the initial mean excitation of the system and $N$ is the asymptotic mean excitation of the system determined by $\gamma$ and $\gamma'$. The condition for the lowest phonon count can be achieved by choosing $\gamma' = X_0^2\gamma$, where $X_0^2 = \tfrac{\hbar}{2m\omega}$ is the variance of the ground state of the harmonic oscillator. 

\subsubsection{General approach to optimising feedback for a particle in a harmonic potential}\label{sec:cooling}
With the intuition gained from the previous calculation, let us consider the expected increments in energy computed according to Eq.~\eqref{eq:da} without approximation:
\begin{align}\label{eq:dH_uv}
\tfrac{1}{\hbar\omega^2}& d\ex{\hat{H}} = \nonumber \\    \Big\{&\tfrac{\kappa}{4} \left[1 - \left(\tfrac{u}{\kappa m\omega^2}\right)^2 - \left(\tfrac{v}{\kappa\omega}\right)^2 \right] \nonumber \\ &  
 +\tfrac{v}{\omega}\left[\tfrac{m\omega}{\hbar}\ex{(\Delta\hat{Q})^2} +\tfrac{1}{2\kappa}\tfrac{v}{\omega} +\tfrac{m\omega}{\hbar}\ex{\hat{Q}}^2\right] \nonumber \\ &
-\tfrac{1}{2}\tfrac{u}{m\omega^2}\left[\tfrac{1}{\hbar}\ex{\{\Delta\hat{Q},\Delta\hat{P}\}} -\tfrac{1}{\kappa}\tfrac{u}{m\omega^2} +\tfrac{2}{\hbar}\ex{\hat{Q}}\ex{\hat{P}}\right]\Big\}dt  \nonumber \\ 
 +&\sqrt{\tfrac{\kappa}{2\omega}}\Big\{2\sqrt{\tfrac{m\omega}{\hbar}}\ex{\hat{Q}} \left[\tfrac{m\omega}{\hbar}\ex{(\Delta\hat{Q})^2} +\tfrac{1}{2\kappa}\tfrac{v}{\omega}\right] \nonumber \\ &  \quad\quad\quad
 +\tfrac{1}{\sqrt{\hbar m\omega}} \ex{\hat{P}} \left[\tfrac{1}{\hbar}\ex{\{\Delta\hat{Q},\Delta\hat{P}\}} -\tfrac{1}{\kappa}\tfrac{u}{m\omega^2}\right] \nonumber \\ &  \quad\quad\quad
  + \left(\tfrac{m\omega}{\hbar}\right)^{\frac{3}{2}} \ex{(\Delta\hat{Q})^3} \nonumber \\ &  \quad\quad\quad
  +\tfrac{1}{2\sqrt{\hbar^3 m \omega}}\ex{\{(\Delta\hat{P})^2,\Delta\hat{Q}\}} \Big\}dW\,.
\end{align}
Here we note that the terms $\tfrac{m\omega}{\hbar}\ex{(\Delta\hat{Q})^2} +\tfrac{1}{2\kappa}\tfrac{v}{\omega}$ and $\tfrac{1}{\hbar}\ex{\{\Delta\hat{Q},\Delta\hat{P}\}} -\tfrac{1}{\kappa}\tfrac{u}{m\omega^2}$  also appear in Eqs.~\eqref{eq:dQ_uv} and \eqref{eq:dP_uv} where they govern the stochastic portion of the increments of the expectation values of the measured observable $\hat{Q}$ and its conjugate momentum $\hat{P}$, respectively.


By choosing the generator of feedback as follows
\begin{align}\label{eq:u}
u & =  \tfrac{m\omega^2 \kappa}{\hbar}\ex{\{\Delta\hat{Q},\Delta\hat{P}\}}_\infty  = m\omega^2\left(- 1 + \sqrt{1+\kappa^2}\right) \\
v & = -\tfrac{2m\omega^2\kappa}{\hbar}\ex{(\Delta\hat{Q})^2}_\infty = -  \sqrt{2}\omega \sqrt{- 1 + \sqrt{1+\kappa^2}} \,,\label{eq:v}
\end{align}
the stochastic terms in Eq.~\eqref{eq:dH_uv}, Eq.~\eqref{eq:dQ_uv} and Eq.~\eqref{eq:dP_uv} are reduced over time, where $\ex{\{\Delta\hat{Q},\Delta\hat{P}\}}_\infty$ and $\ex{(\Delta\hat{Q})^2}_\infty$ are given by Eq.~\eqref{eq:anti_com_inf_kappa} and Eq.~\eqref{eq:Q2_inf_kappa}, respectively. We note that for weak measurement, $\kappa \ll 1$, this choice of feedback coincides with our earlier result obtained for an optical oscillator, namely $u = 0$ and $v  = - \kappa \omega$.  

\begin{figure}
     \centering
     \begin{subfigure}[b]{0.45\textwidth}
         \centering
     	\includegraphics[width=\textwidth]{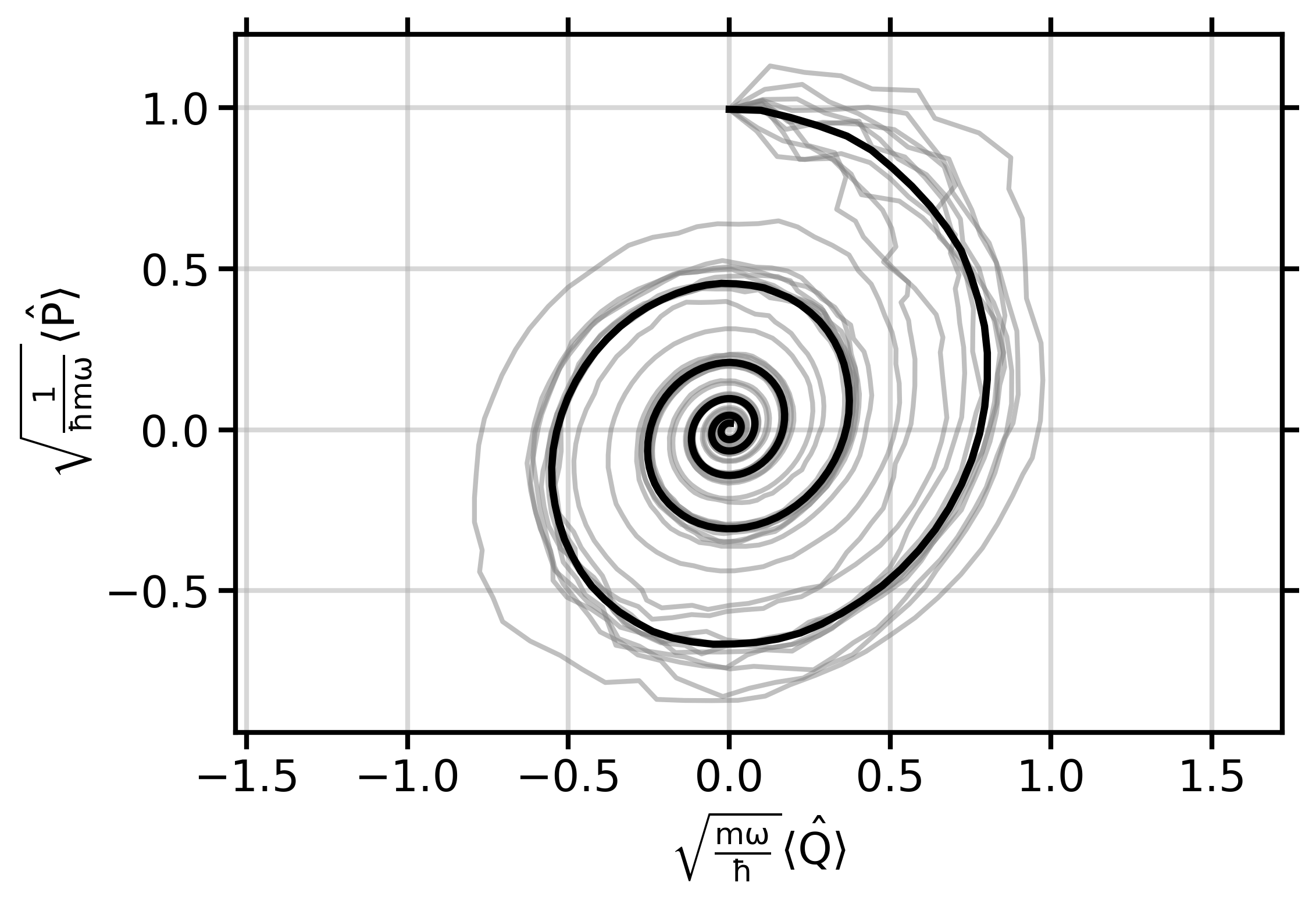}
     	\caption{Time evolution of the unitless expectation values of the measured observable $M = Q$ and the conjugate momentum $P$. }\label{fig:23}
     \end{subfigure}
     \hfill
     \begin{subfigure}[b]{0.45\textwidth}
         \centering
    	\includegraphics[width=\textwidth]{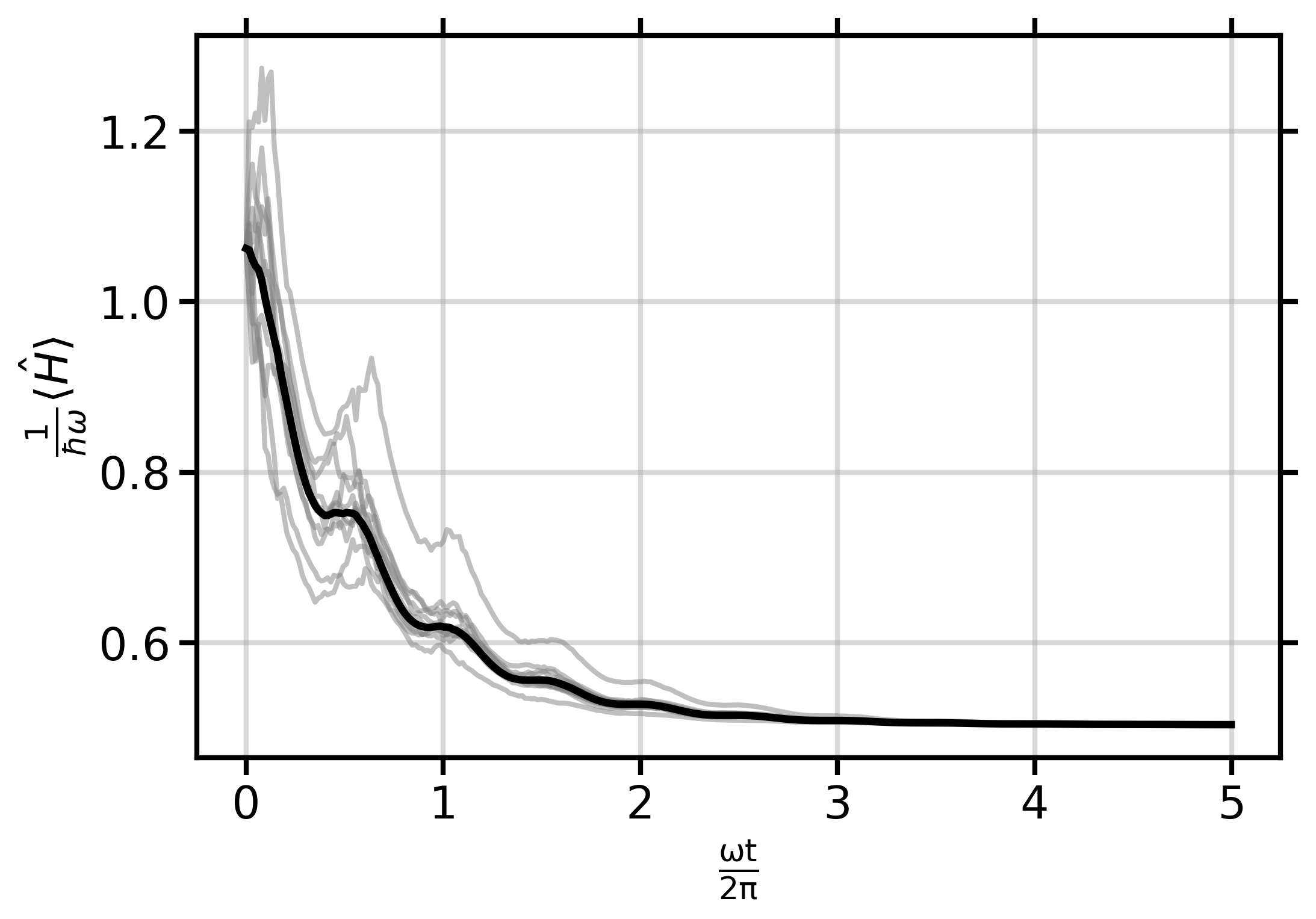}
     	\caption{Time evolution of the unitless version of the expected energy.}\label{fig:25}
    \end{subfigure}
     \caption{The following results were obtains by numerical simulation using the second-order weak scheme~\cite{breuer2002theory} with feedback chosen according to Eqs.~\eqref{eq:u} and \eqref{eq:v} and constant $\kappa=0.25$. The initial conditions were $\ex{\hat{Q}} = 0$ and $\ex{\hat{P}} = 1$ and those of the second-order variances are given in Figure~\ref{fig:21}. Grey lines indicates individual simulations while the black line indicates the mean values taken over 100 simulations.}  
\end{figure}

After some time $t\gg \tfrac{1}{\omega\kappa}$ the stochastic terms in  the increments of the expectation value of the measured observable \eqref{eq:dQ_uv} and  its conjugate momentum \eqref{eq:dP_uv} are negligible, so that
\begin{align}
\frac{d\ex{\hat{Q}}}{dt} = & \tfrac{1}{m}\ex{\hat{P}} -  \sqrt{2}\omega \sqrt{- 1 + \sqrt{1+\kappa^2}}\ex{\hat{Q}} \\
\frac{d\ex{\hat{P}}}{dt} = & -m\omega^2\sqrt{1+\kappa^2}\ex{Q}\,.
\end{align}
The above equations are equivalent to the equations of motion of a damped harmonic oscillator and solving them yields a decay rate for the expectation values of $Q$ and $P$ of
\begin{align}
\frac{\omega}{\sqrt{2}} \sqrt{-1+\sqrt{1+\kappa^2}}\,.
\end{align}
For $\kappa<< 1$, the decay rate is approximately $\kappa\omega/2$. In Figure~\ref{fig:23} the results of a numerical simulation of the time evolution of the expectation values of the measured observable and its conjugate momentum are shown for $\kappa = 0.25$, under the assumption that the initial wavefunction of the system is Gaussian. 

For times $t >> \frac{1}{\omega\kappa}$, the increment in the expectation value of the energy~\eqref{eq:dH_uv} is given by
\begin{align}
\tfrac{1}{\hbar\omega^2}d\ex{\hat{H}} = & \left( \tfrac{v}{\omega}\tfrac{m\omega}{\hbar}\ex{\hat{Q}}^2 
-\tfrac{u}{m\omega^2}\tfrac{1}{\hbar}\ex{\hat{Q}}\ex{\hat{P}}\right) dt\,.
\end{align}
The energy, as well as the increment in the energy, decays to its asymptotic value at a rate twice that of the decay of the measured observable $Q$ and the conjugate momentum $P$. For weak measurement $\kappa << 1$, this decay rate coincides with the decay rate $\gamma'= \omega\kappa$ found for an optical oscillator in the rotating wave approximation and the non-selective regime of measurement. In Figure~\ref{fig:25} a numerical simulation of the evolution of the energy over time illustrates this exponential decay.

For feedback determined by Eqs.~\eqref{eq:u} and \eqref{eq:v}, $\ex{\hat{Q}}_\infty$ and $\ex{\hat{P}}_\infty$  are negligible at times $t>>\tfrac{1}{\omega\kappa}$, and the particle is approximately stationary in the centre of the potential with an asymptotic energy given by 
\begin{align}
E_\infty & = \frac{\hbar \omega}{2\sqrt{2}}\frac{\kappa}{\sqrt{- 1 + \sqrt{1+\kappa^2}}}\,.
\end{align}
As discussed in Section~\ref{sec:stat_har}, the energy of the particle converges to the ground state energy with respect to the co-moving frame in the weak measurement limit. We can therefore combine weak measurement and feedback to achieve confinement, resulting in a convergence to the ground state in the lab frame.


\section{Discussion}
In this paper we investigated continuous measurement and feedback on a particle in a harmonic potential, where the observable and the generator of feedback are arbitrary linear combinations of position and momentum. We have shown that the asymptotic state, assumed by the particle after a certain convergence time determined by the measurement strength, is represented by a Gaussian wave function. Its width does not depend on the feedback but only on the product of the measurement strength, the variance of the ground state and the time period of the oscillator. On the other hand, the location of the Gaussian in phase space depends on the feedback. This is not surprising, since the feedback merely shifts the wave function in position and/or momentum space. 

\begin{figure}
 \centering
     \begin{subfigure}[b]{0.45\textwidth}
         \centering
     	\includegraphics[width=\textwidth]{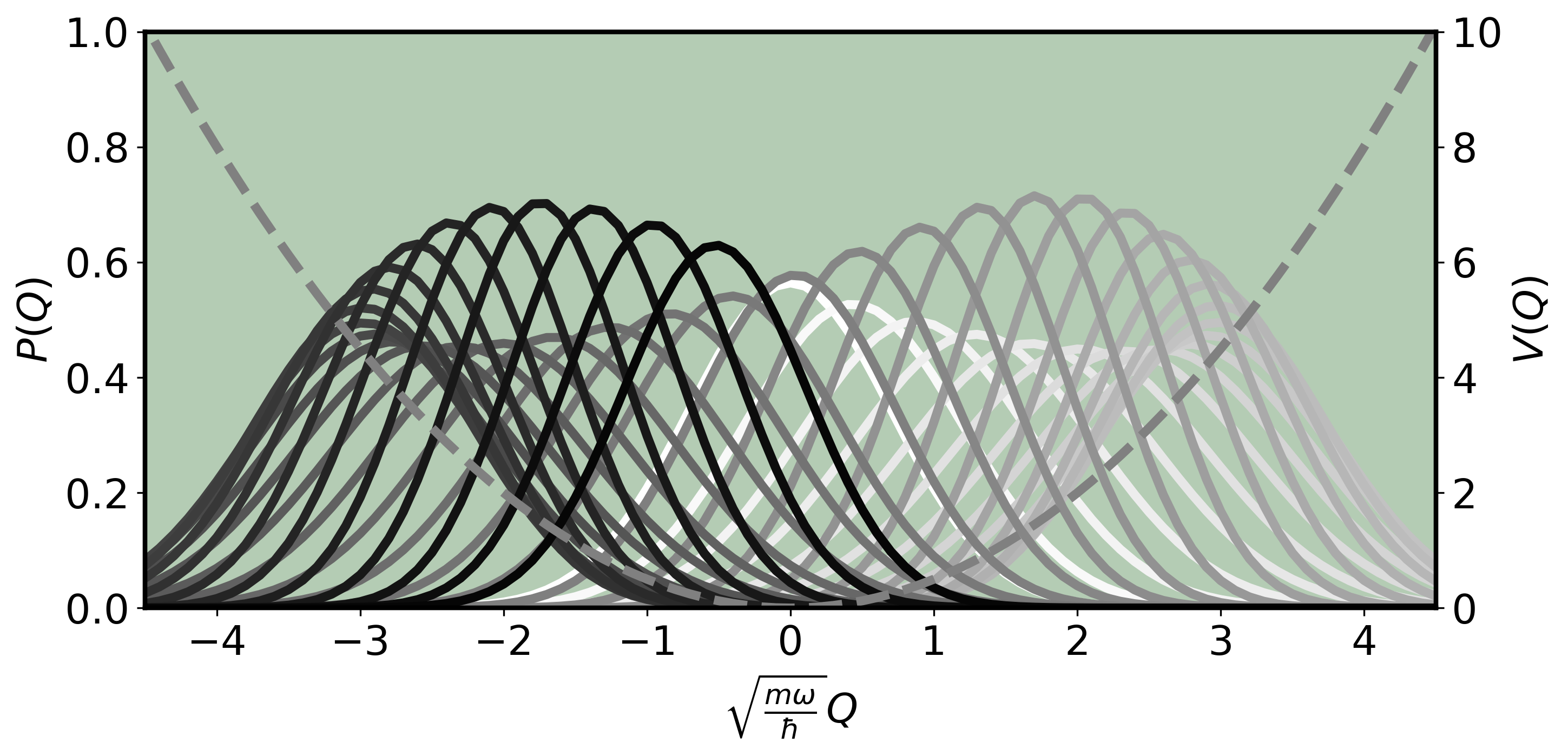}
     	\caption{Evolution of the wave function in $Q$-space in the absence of measurement. }\label{fig:harmonic}
     \end{subfigure}
     \hfill
     \begin{subfigure}[b]{0.45\textwidth}
         \centering
     	\includegraphics[width=\textwidth]{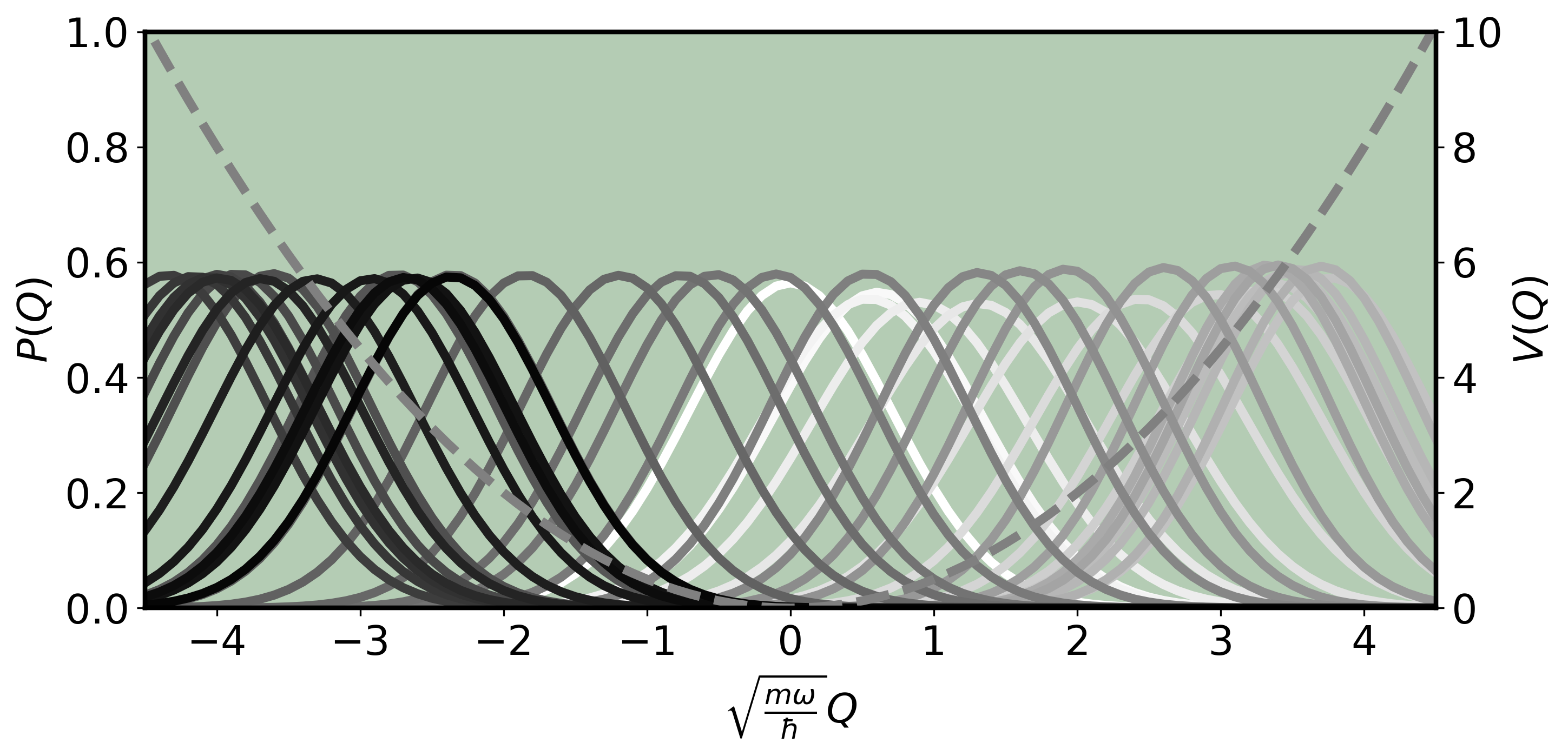}
     	\caption{Measurement introduces noise into the dynamics of the system. In addition, the width of the wave function stabilises over time. }\label{fig:harmonic+m}
     \end{subfigure}
     \hfill
     \begin{subfigure}[b]{0.45\textwidth}
         \centering
     	\includegraphics[width=\textwidth]{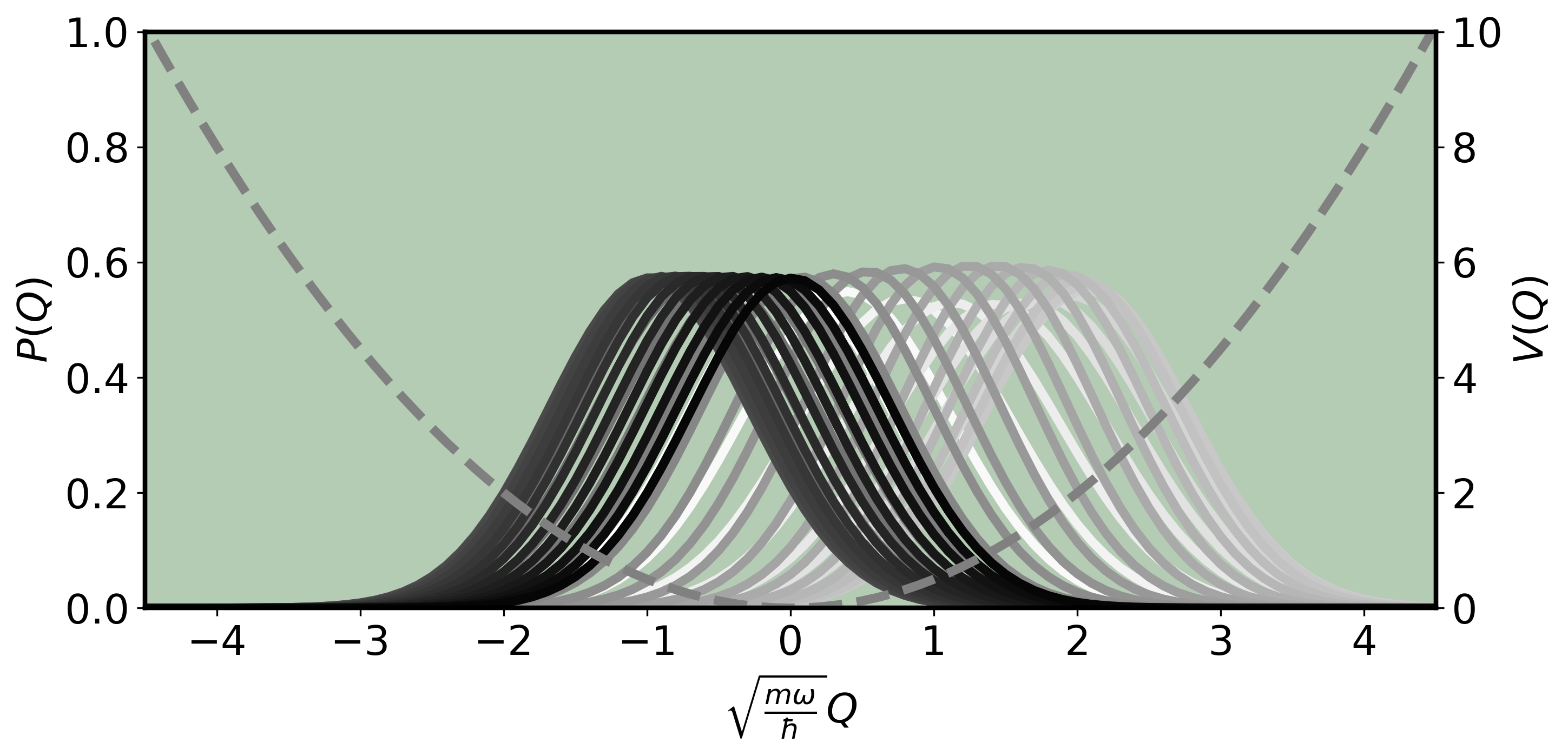}
     	\caption{By measurement and appropriate feedback, the wave function is localised in the centre of the potential with a stable width. }\label{fig:harmonic+m+f}
     \end{subfigure}
         \caption{Comparison of dynamics in the absence of measurement (a), with measurement (b) and with measurement and feedback (c). The initial state of the wave function is shown in white, while the evolved wave function is indicated in increasingly darker shades of grey. A dotted line represents the shape of the potential $V(Q) = \tfrac{m\omega^2}{2}Q^2$, with unitless values shown on the right axis. The probability $P(Q)$ per interval of length $dQ$ is given on the left axis.}
\label{fig:comp}
\end{figure}

It is possible to localise the wave function in the centre of the harmonic potential with vanishing momentum expectation value. The harmonic oscillator wave function for the different scenarios (with and without measurement and feedback) is depicted in Figure~\ref{fig:comp}. Feedback enables localisation in the ground state, measurement allows the wave function to climb up the potential mountain without being squeezed as is the case in the absence of measurement. This surprising effect can be understood in the Heisenberg Picture where the observable is rotated periodically into its conjugate momentum, resulting in a homogeneous localisation of the particle. 

Our analysis shows that, remarkably, for a free particle or a particle in a harmonic potential, the feedback can be employed to cancel the measurement noise completely, cp.~Eqs.~\eqref{eq:dQ_uv} and \eqref{eq:dP_uv}. This can be achieved by adjusting the feedback to the values of the variances of the measured observable and its conjugate momentum. These values can be obtained analytically if the initial state is a known Gaussian state. For arbitrary initial states the values are known in the asymptotic regime as a function of the relative measurement strength. They can also be eventually inferred from the measurement results (cp.~state monitoring~\cite{konrad2010monitoring}).
By combining this confining feedback with sufficiently weak measurement, it is possible to cool the system to the ground state.

We conjecture that cooling is also possible for arbitrary potentials. Feedback compensating the noise that contributes to the change of the expectation value of the measured observable simultaneously damps the motion of the particle. For sufficiently strong measurement, the asymptotic wave function can be localised on a length scale such that the potential can be approximated by a polynomial of second-order\cite{halliwell1995quantum}. Thus the dynamics is reduced to the case of the harmonic oscillator as discussed above. 

On the other hand, it is also possible to control the motion of the particle by means of feedback which adds an extra term quadratic in the measured observable $Q$ to the Hamiltonian. This allows for modifying or even compensating the kinetic or potential energy in the Hamiltonian. However, such dynamic control contributes in general to the generation of noise, since it is not compatible with the compensation of noise. 

\section*{Acknowledgements}

Amy R thanks the National Research Foundation of South Africa for funding. Anirudh R, SM and TK acknowledge funding by the South African Quantum Technology Initiative grant from the Department of Science and Innovation of South Africa.

\appendix
\section{Ito Equations}\label{ap:ito}

In this appendix we derive the feedback master equation for an applied feedback corresponding to the action of the unitary operator given in Eq.~\eqref{eq:unitary_operator}. The unitary that describes the feedback is given by
\begin{align}
  U &= \exp\left[-\frac{i}{\hbar}\hat{F} d\mathcal{M}\right] \nonumber \\
    &= \mathbb{I} - \frac{i}{\hbar}\hat{F}d\mathcal{M} - \frac{1}{2\hbar^2}\hat{F}^2d\mathcal{M}^2\nonumber \\
    &= \mathbb{I} - \frac{i}{\hbar} \hat{F} d\mathcal{M} - \frac{1}{2\hbar^2 \gamma}\hat{F}^2dt\,,
\end{align}
where the Ito rules, Eq.~\eqref{eq:Ito_rules} are used and all terms $\mathcal{O}(dt^2)$ are neglected. Application of the unitary feedback to the state of the system after measurement leads to
\begin{align}
  U&(\rho + d\rho) U^\dagger \nonumber \\
  = &\rho + d\rho - \frac{i}{\hbar}\left[\hat{F},\rho\right]d\mathcal{M} - \frac{i}{\hbar}\left[\hat{F},d\rho\right]d\mathcal{M} \nonumber \\ &-\frac{1}{2\hbar^2\gamma}\left[\hat{F},\left[\hat{F},\rho\right]\right]dt \nonumber \\
  =& \rho + d\rho - \frac{i}{\hbar}\left[\hat{F},\rho\right]\left(\langle \hat{M} \rangle dt + \tfrac{1}{\sqrt{\gamma}} dW\right)\nonumber \\ & - \frac{i}{\hbar\sqrt{\gamma}}\left[\hat{F},d\rho\right] dW-\frac{1}{2\hbar^2 \gamma}\left[\hat{F},\left[\hat{F},\rho\right]\right]dt\,. \label{eq:calc}
\end{align}
We can now substitute the expression for $d\rho$ given by the Ito differential equation~\eqref{eq:ito_dif}, into the third term of the Eq.~(\ref{eq:calc}) such that
\begin{align}
 \frac{i}{\sqrt{\gamma}} \left[\hat{F},d\rho\right] dW 
=& \frac{i}{\sqrt{\gamma}}\left[\hat{F}, \tfrac{\sqrt{\gamma}}{2}\left\{\hat{M} - \langle\hat{M}\rangle,\rho\right\}\right]dt  \nonumber \\
  =& \frac{i}{2}\left[\hat{F}, \left\{\hat{M},\rho\right\} - \left\{\langle \hat{M}\rangle,\rho\right\}\right]dt \nonumber \\
  =&\frac{i}{2}\left[\hat{F}, \left\{\hat{M},\rho\right\}\right]dt + i\langle \hat{M}\rangle\left[\hat{F},\rho\right]dt\,.
\end{align}
Hence the state of the system after the application of feedback is given by
\begin{align}
 U(\rho + d\rho) U^\dagger 
 = & \rho + d\rho  - \frac{i}{2\hbar}\left[\hat{F}, \left\{\hat{M},\rho\right\}\right]dt \nonumber \\ 
&  -\frac{1}{2\hbar \gamma}\left[\hat{F},\left[\hat{F},\rho\right]\right]dt - \frac{i}{\hbar \sqrt{\gamma}}\left[\hat{F},\rho\right]dW  \nonumber \\
= & \rho + d\rho'\,,
\end{align}
where
\begin{align} \label{eq:gen_feedback}
  d\rho' :=& -\frac{i}{\hbar}\left[\hat{H},\rho\right]dt - \frac{\gamma}{8}\left[\hat{M},\left[\hat{M},\rho\right]\right]dt  \nonumber \\
  &-\frac{1}{2\hbar^2 \gamma}\left[\hat{F},\left[\hat{F},\rho\right]\right]dt -\frac{i}{2\hbar }\left[\hat{F}, \left\{\hat{M},\rho\right\}\right]dt   \nonumber \\
  &+ \frac{\sqrt{\gamma}}{2}\left\{\hat{M} - \langle \hat{M} \rangle,\rho\right\}dW - \frac{i}{\hbar \sqrt{\gamma}}\left[\hat{F},\rho\right]dW\,.
\end{align}

\section{Equation of motion of a pure state in the co-moving reference frame}\label{app:co-moving}

In order to determine the state $\ket{\tilde{\psi}}+d\ket{\tilde{\psi}}$ the displacement operators in Eq.~\eqref{eq:displacement} are first combined into a single exponential,
\begin{align}
\exp\big\{-\tfrac{i}{\hbar} d\ex{\hat{p}} \hat{q}\big\} &\exp\big\{\tfrac{i}{\hbar} d\ex{\hat{q}} \hat{p}\big\} \nonumber\\ &\quad = \exp\big\{-\tfrac{i}{\hbar} \left( d\ex{\hat{p}} \hat{q} - d\ex{\hat{q}} \hat{p}\right)  \big\}\,,
\end{align}
where we ignore the irrelevant phase factor which results from applying the  Baker–Campbell–Hausdorff formula. Eq.~\eqref{eq:psi_dpsi} can now be written as
\begin{align}
\ket{\tilde{\psi}}+d\ket{\tilde{\psi}} &  = e^{-\tfrac{i}{\hbar} \left( d\ex{\hat{p}} \hat{q} - d\ex{\hat{q}} \hat{p}\right)}e^{\hat{G}} (\ket{{\psi}}+d\ket{{\psi}})\,,
\end{align}
where $\hat{G}$ is given in Eq.~\eqref{eq:G}.  As noted previously $\hat{q}$ and $\hat{p}$ can be replaced with any pair of conjugate variables. We therefore replace them with $\hat{Q}$ and $\hat{P}$, defined in Eqs.~\eqref{eq:Q} and \eqref{eq:P_alpha_beta}, so that
\begin{align}
\ket{\tilde{\psi}}+d\ket{\tilde{\psi}} &  = e^{-\tfrac{i}{\hbar} \left( d\ex{\hat{P}} \hat{Q} - d\ex{\hat{Q}} \hat{P}\right)}e^{\hat{G}} (\ket{{\psi}}+d\ket{{\psi}})\,,\label{ap:psi_tilde}
\end{align}

Assuming that $\ex{\hat{Q}} =0 = \ex{\hat{P}}$ at time $t$ when $\ket{\psi}$ is considered, we have
\begin{align}
\hat{G} =& \Big[-\tfrac{i}{\hbar}\hat{H}  -\tfrac{\gamma}{4}\hat{Q}^2  + \tfrac{i}{4\hbar} [\hat{Q},\hat{F}] \Big] dt \nonumber \\ &+\Big[\tfrac{\sqrt{\gamma}}{2}\hat{Q} -\tfrac{i}{\hbar \sqrt{\gamma}}\hat{F} \Big]dW\,.
\end{align}
and
\begin{align}
d\ex{Q} & = \operatorname{Tr}[{\hat{Q} d\rho}] \nonumber \\ &= -\tfrac{i}{\hbar}\ex{[\hat{Q},\hat{H}]}dt -\tfrac{1}{2\hbar^2\gamma}\ex{[\hat{F},[\hat{F},\hat{Q}]]}dt \nonumber \\ &\quad
-\frac{i}{2\hbar}\ex{\{\hat{Q},[\hat{Q}, \hat{F}]\}}dt \nonumber \\ &\quad + \frac{\sqrt{\gamma}}{2} \ex{\{\hat{Q},\hat{Q}\}} dW   + \frac{i}{\hbar\sqrt{\gamma}}\ex{[\hat{F},\hat{Q}]} dW\,.
\end{align}
Given that feedback is a linear combination of position and momentum,  $[\hat{F},\hat{Q}]$ and $[\hat{F},\hat{P}]$ are constant, so that
\begin{align}
d\ex{Q}  =& -\tfrac{i}{\hbar}\ex{[\hat{Q},\hat{H}]}dt \nonumber \\ & + \frac{\sqrt{\gamma}}{2} \ex{\{\hat{Q},\hat{Q}\}} dW   + \frac{i}{\hbar\sqrt{\gamma}}\ex{[\hat{F},\hat{Q}]} dW\,. \label{ap:dq}
\end{align}
Similarly,
\begin{align}
d\ex{P} & = \operatorname{Tr}[{\hat{P}d\rho}] \nonumber \\ &= -\tfrac{i}{\hbar}\ex{[\hat{P},\hat{H}]}dt -\frac{1}{2\hbar^2\gamma}\ex{[\hat{F},[\hat{F},\hat{P}]]}dt \nonumber \\ &\quad
-\frac{i}{2\hbar}\ex{\{\hat{Q},[\hat{P}, \hat{F}]\}}dt \nonumber \\ &\quad + \frac{\sqrt{\gamma}}{2} \ex{\{\hat{Q},\hat{P}\}} dW + \frac{i}{\hbar\sqrt{\gamma}}\ex{[\hat{F},\hat{P}]} dW \nonumber \\ 
&=-\tfrac{i}{\hbar}\ex{[\hat{P},\hat{H}]}dt \nonumber \\ &\quad + \frac{\sqrt{\gamma}}{2} \ex{\{\hat{Q},\hat{P}\}} dW + \frac{i}{\hbar\sqrt{\gamma}}\ex{[\hat{F},\hat{P}]} dW\,.
\label{ap:dp}
\end{align}

The Baker–Campbell–Hausdorff formula can now be applied to Eq.~\eqref{ap:psi_tilde}. Hence,
\begin{align}
&\ket{\tilde{\psi}} + d\ket{\tilde{\psi}} \nonumber \\  
&=\exp\Bigg\{ -\tfrac{i}{\hbar}\left(\hat{{H}}-\tfrac{i}{\hbar}\ex{[\hat{P},\hat{H}]} \hat{Q} +\tfrac{i}{\hbar}\ex{[\hat{Q},\hat{H}]}\hat{P}\right)dt \nonumber \\ &\hspace{40pt} -\tfrac{\gamma}{4}\hat{Q}^2dt +\frac{i}{4\hbar} [\hat{Q},\hat{F}]dt + \hat{A}  dt\nonumber\\ & \hspace{17pt}-\tfrac{i}{\hbar}\sqrt{\tfrac{\gamma}{2}} \left[ \left(i\hbar + \ex{\{\hat{Q},\hat{P}\}} \right)\hat{Q} - 2\ex{\hat{Q}^2} \hat{P}\right] dW \Bigg\} \ket{\tilde{\psi}}\label{app_eq:psi_dpsi} \,,
\end{align}
where we use the fact that $\ex{[\hat{F},\hat{P}]}\hat{Q}-\ex{[\hat{F},\hat{Q}]}\hat{P} = i\hbar \hat{F}$. $\hat{A}$ is the term which arises from the commutation relation between the arguments of the two exponential terms in Eq.~\eqref{ap:psi_tilde}. Computing the term $\hat{A}dt$ yields
\begin{align}
\hat{A} dt   = &-\tfrac{i}{2\hbar} \left[ d\ex{\hat{P}} \hat{Q} - d\ex{\hat{Q}} \hat{P} ,\hat{G}\right]
 \nonumber\\  =& \tfrac{i}{2\hbar}  \left[ -d\ex{\hat{P}} \hat{Q}  +d\ex{\hat{Q}} \hat{P},\tfrac{\sqrt{\gamma}}{2}\hat{Q} -\tfrac{i}{\hbar\sqrt{\gamma}}\hat{F} \right] dW \nonumber\\ 
=  &  \tfrac{\sqrt{\gamma}}{4} d\ex{\hat{Q}} dW -\tfrac{1}{2\hbar^2} \left( d\ex{\hat{P}} [\hat{Q},\hat{F}] - d\ex{\hat{Q}} [\hat{P},\hat{F}]\right) dW \nonumber\\  
=  &  \tfrac{\sqrt{\gamma}}{4} d\ex{\hat{Q}} dW  -\tfrac{i\sqrt{\gamma}}{4\hbar^2} \ex{\{\hat{Q},\hat{F}\}} dt \,,
\end{align}
where we have used the fact that $\hat{F}$ is a linear combination of $\hat{Q}$ and $\hat{P}$ to obtain the last  result. Substituting Eq.~\eqref{ap:dq} into the above gives
\begin{align}
\hat{A} = & \tfrac{\gamma}{4} \ex{\hat{Q}^2} +\tfrac{i}{4\hbar} \ex{[\hat{F},\hat{Q}]} 
 + \tfrac{i}{4\hbar} \ex{\{\hat{Q},\hat{F}\}}\,. \label{app_eq:A}
\end{align}
Substituting the above into Eq.~\eqref{app_eq:psi_dpsi} while ignoring the phase factor associated with the last term in Eq.~\eqref{app_eq:A}, leads to the expression
\begin{widetext}
\begin{align}
 \ket{\tilde{\psi}} + d\ket{\tilde{\psi}}  =  \exp\Big\{ \Big[-\tfrac{i}{\hbar}\left(\hat{{H}}-\tfrac{i}{\hbar}\ex{[\hat{P},\hat{H}]} \hat{Q} +\tfrac{i}{\hbar}\ex{[\hat{Q},\hat{H}]}\hat{P}\right) 
    -\tfrac{\gamma}{4}\left(\hat{Q}^2 - \ex{\hat{Q}^2}\right) \Big]& dt \nonumber \\ 
    -\tfrac{i}{\hbar}\sqrt{\tfrac{\gamma}{2}} \Big[ \left(i\hbar + \ex{\{\hat{Q},\hat{P}\}} \right)\hat{Q} 
    - 2\ex{\hat{Q}^2} \hat{P}\Big]& dW 
    \Big\} \ket{\tilde{\psi}}\,. \label{ap:state_change_genH}
\end{align}
\end{widetext}
If $\hat{H}$ is the Hamiltonian of a harmonic oscillator \eqref{eq:H_QP} then $\ex{[\hat{Q},\hat{H}]}=\ex{[\hat{P},\hat{H}]}=0$, since $\ex{\hat{Q}}=\ex{\hat{P}} = 0$, and 
\begin{align}
 &  \ket{\tilde{\psi}} + d\ket{\tilde{\psi}} \nonumber \\  &=\exp\Big\{ \Big[-\tfrac{i}{\hbar}\hat{{H}} -\tfrac{\gamma}{4}\left(\hat{Q}^2 - \ex{\hat{Q}^2}\right) \Big] dt
 \nonumber\\ &   \hspace{25pt}
 -\tfrac{i}{\hbar}\sqrt{\tfrac{\gamma}{2}} \Big[ \left(i\hbar + \ex{\{\hat{Q},\hat{P}\}} \right)\hat{Q} - 2\ex{\hat{Q}^2} \hat{P}\Big] dW \Big\} \ket{\tilde{\psi}}\,.
\end{align}

\section{Proof that the Gaussian steady-state solutions satisfy the energy eigenvalue equation}\label{ap:eigen}

This appendix will show that the stationary solution to the eigenvalue equation~\eqref{eq:terms_dW} also satisfies the energy eigenvalue equation~\eqref{eq:terms_dt}. The eigenvalue equation~\eqref{eq:terms_dW}, where $[\hat{Q},\hat{P}]=i\hbar$, has in ``$Q$" representation the solution
\begin{align}
\psi_\infty(Q) = \left(\frac{s}{\pi}\right)^{\frac{1}{4}} \exp\left({- \frac{s}{2} u^2}\right)\label{app:gaussian_Q}
\end{align}
where 
\begin{align}
s : = \frac{1}{2\ex{(\hat{Q})^2}_\infty } \left(1 -\frac{i}{\hbar }\ex{\{\hat{Q},\hat{P}\}}_\infty \right) \,.
\end{align}

 The energy eigenvalue equation~\eqref{eq:terms_dt} can also be expressed in terms of $\hat{Q}$ and $\hat{P}$ as
\begin{align}
\left[ \frac{1}{2m}\hat{P}^2+\frac{m\omega^2}{2}\hat{Q}^2  -i \tfrac{\hbar \gamma}{4}\left(\hat{Q}^2 - \ex{\hat{Q}^2}_\infty\right)\right] & \ket{\tilde{\psi}_\infty} \nonumber \\= E &\ket{\tilde{\psi}_\infty}  \,.\label{ap:eigen_dt}
\end{align}
Using Eq.~\eqref{app:gaussian_Q} the energy eigenvalues equation~\eqref{ap:eigen_dt} yields
\begin{align}
E   =& \frac{\hbar^2 s}{2m} +i \frac{\hbar \gamma}{4} \ex{(\hat{Q})^2}_\infty \nonumber \\ & + \frac{m\omega^2}{2}\left(1 - \left(\frac{\hbar s}{m\omega}\right)^2  -i \frac{\hbar \gamma}{2m\omega^2}  \right)Q^2   \,.\label{ap:dummy_E}
\end{align}
It is possible to evaluate the term proportional to $Q^2$ using the result given in Eqs.~\eqref{eq:anti_com_inf_kappa} and \eqref{eq:Q2_inf_kappa}. A straight forward calculation gives
\begin{align}
\left(\frac{\hbar s}{m\omega}\right)^2& = 1-i\kappa = 1- i\frac{\hbar \gamma}{2m\omega^2} \,,
\end{align}
where $\kappa$ is defined in Eqn.~\eqref{eq:kappa_dum}. The $Q^2$ term therefore vanishes and we can conclude from Eq.~\eqref{ap:dummy_E} that
\begin{align}
E &= \frac{\hbar^2}{4m}\left(2s +i \frac{ m\gamma}{\hbar} \ex{(\hat{Q})^2}_\infty\right)\nonumber \\ &  = \frac{\hbar^2 }{4m \ex{(\hat{Q})^2}_\infty}\\
& = \frac{\hbar \omega}{2\sqrt{2}}\frac{\kappa}{\sqrt{- 1 + \sqrt{1+\kappa^2}}}  \,.
\end{align}
For for a free particle, i.e.~$\omega\rightarrow 0$, this solution equivalent to the energy determined by Di\'{o}si in \cite{diosi1988localized}. In addition, for $\kappa>>1$,
\begin{align}
E \approx  \frac{\hbar\omega}{2\sqrt{2}} \sqrt{\kappa}\,,
\end{align}
and for $\kappa<<1$, 
\begin{align}
E \approx  \frac{\hbar\omega}{2} \left(1+\frac{\kappa^2}{8}\right)\,.
\end{align}
In the weak measurement limit, $\kappa\rightarrow 0$, the energy approaches  the value associated with the ground state energy of the harmonic oscillator, $\hbar\omega/2$.

\section{Dimensionless coupled differential equations}\label{Appendix:dimensionless}
In this section we show that by dividing Eqs.~\eqref{eq:td_1}-\eqref{eq:td_3} by the appropriate constants of matching dimension it is possible to greatly simplify these coupled differential equations, as shown below.
\begin{widetext}
\begin{align}
\tfrac{m\omega}{\hbar}d\ex{ (\Delta \hat{Q})^2} = & -2 \kappa\left(\tfrac{m\omega}{\hbar}\right)^2\ex{ (\Delta \hat{Q})^2}^2 \omega dt +\tfrac{1}{\hbar} \ex{\{\Delta \hat{P}, \Delta \hat{Q} \}} \omega dt +\sqrt{2\kappa}\left(\tfrac{m\omega}{\hbar}\right)^{\frac{3}{2}} \ex{(\Delta \hat{Q})^3} \sqrt{\omega}dW\label{ap:td_1} 
\\
\tfrac{1}{\hbar m \omega}d\ex{ (\Delta \hat{P})^2}=  & \tfrac{1}{2} \kappa\left( 1 - \tfrac{1}{\hbar^2}\ex{\{\Delta \hat{P}, \Delta \hat{Q} \}}^2\right) \omega dt   -\tfrac{1}{\hbar}\ex{\{\Delta \hat{P}, \Delta \hat{Q} \}}  \omega dt  +\sqrt{\tfrac{\kappa}{2 }}\tfrac{1}{\sqrt{\hbar^3 m\omega}} \ex{\{ (\Delta \hat{P})^2, \Delta \hat{Q}\}} \sqrt{\omega}dW \label{ap:td_2}
\\
\tfrac{1}{\hbar} d\ex{\{\Delta \hat{Q}, \Delta \hat{P} \}} = & 2 \left(\tfrac{1}{\hbar m\omega} \ex{ (\Delta \hat{P})^2}  - \tfrac{m\omega}{\hbar}   \ex{ (\Delta \hat{Q})^2}\right) \omega dt  -  2\kappa \tfrac{m\omega}{\hbar^2} \ex{\{\Delta \hat{P}, \Delta \hat{Q} \}}\ex{ (\Delta \hat{Q})^2} \omega dt   \nonumber \\ & + \sqrt{\tfrac{\kappa}{2}}\sqrt{\tfrac{m \omega}{\hbar^3}} \ex{\{ \{\Delta \hat{Q}, \Delta \hat{P} \}, \Delta \hat{Q} \}} \sqrt{\omega} dW \,.\label{ap:td_3}
\end{align}
\end{widetext}
For a free particle or a particle in a harmonic potential and for $\tau = \omega t >> {1}/{\kappa\omega}$, higher order moments can be ignored. By substitution the variables defined in Eqs.~\eqref{eq:x}-\eqref{eq:z} we arrive at the simple equations
\begin{align}
\frac{dx}{d\tau}& = -2\kappa x^2 + z \label{ap:dx}\\
\frac{dy}{d\tau}& = \frac{\kappa}{2}\left(1-z^2\right)-z\label{ap:dy}\\
\frac{dz}{d\tau}& = 2\left(y-x\right)-2\kappa x z\label{ap:dz}\,.
\end{align}

\section{General feedback equation for Harmonic oscillator in terms of creation and annihilation operators} \label{AppendixC}

According to the definitions of the creation and annihilation operators given in Eq.~\eqref{eq:newqp}, the measurement and feedback operators can be expressed, respectively, as
\begin{align}
\hat{M} & =  c_1^*\hat{a} + c_1\hat{a}^\dagger\label{app:M}\\
\hat{F} & = c_2^*\hat{a} + c_2\hat{a}^\dagger\label{app:F}
\end{align}
where 
\begin{align}
c_1 & = \alpha\sqrt{\tfrac{\hbar}{m\omega}}+i\beta\sqrt{\hbar m\omega}\\
c_2 & = \chi\sqrt{\tfrac{\hbar}{m\omega}}+i\delta\sqrt{\hbar m\omega}\,.
\end{align}
Substituting Eqs.~\eqref{app:M} and \eqref{app:F} into Eq.~\eqref{eq:stochastic_master} we obtain
\begin{align}\label{eq:gen_creation}
  d\rho' =& -\tfrac{i}{\hbar}[\hat{H},\rho]  \nonumber \\
  &- \tfrac{\gamma}{16}\left[c_1^*\hat{a} + c_1\hat{a}^\dagger,\left[c_1^*\hat{a} + c_1\hat{a}^\dagger,\rho\right]\right]dt  \nonumber \\
  & -\tfrac{1}{4\gamma\hbar^2}[c_2^*\hat{a} + c_2\hat{a}^\dagger,\left[c_2^*\hat{a} + c_2 \hat{a}^\dagger,\rho\right]]dt \nonumber \\
  &- \tfrac{i}{4\hbar}\left[c_2^*\hat{a} + c_2 \hat{a}^\dagger, \left\{c_1^*\hat{a}+ c_1\hat{a}^\dagger,\rho\right\}\right]dt\nonumber \\
  & - \tfrac{i}{\sqrt{2\gamma}\hbar}\left[c_2^*\hat{a} + c_2 \hat{a}^\dagger,\rho\right]dW \nonumber \\
  &+ \tfrac{\sqrt{\gamma}}{2\sqrt{2}}\left\{c_1^* \hat{a}+ c_1\hat{a}^\dagger - \langle c_1^*\hat{a}+ c_1^* \hat{a}^\dagger\rangle,\rho\right\}dW \,.
\end{align}
The double commutator $[c_i^*\hat{a} + c_i\hat{a}^\dagger,\left[c_i^*\hat{a} + c_i \hat{a}^\dagger,\rho\right]]$ can be written as
\begin{align}\label{app:double_com}
&c_i^{*2}\left(\{\hat{a}^2,\rho\} - 2 \hat{a}\rho\hat{a}\right)+ c_i^2\big(\{\hat{a}^{\dagger^2},\rho\} - 2\hat{a}^\dagger\rho\hat{a}^\dagger\big) \nonumber \\
& + |c_i|^2(\{\hat{a}^\dagger\hat{a},\rho\} - 2\hat{a}^\dagger\rho\hat{a} + \{\hat{a}\hat{a}^\dagger,\rho\} - 2\hat{a}\rho\hat{a}^\dagger ) \,.
\end{align}
Similarly, the mixed term $\left[c_2^*\hat{a} + c_2 \hat{a}^\dagger, \left\{c_1^*\hat{a}+ c_1\hat{a}^\dagger,\rho\right\}\right]$ can be written as
\begin{align}
&c_1c_2 \left[\hat{a}^{\dagger2},\rho\right] +c _1^*c_2^* \left[\hat{a}^{2},\rho\right]\nonumber \\
&+ c_1c_2^* \left(\hat{a}\hat{a}^\dagger \rho   -\rho \hat{a}^\dagger\hat{a}  \right) + c_1^*c_2 \left(\hat{a}^\dagger\hat{a} \rho   -\rho \hat{a}\hat{a}^\dagger \right)\nonumber \\
& + \left(c_1c_2^* - c_1^*c_2\right) \left(\hat{a}\rho\hat{a}^\dagger  - \hat{a}^\dagger\rho\hat{a}\right)\,.
\end{align}
\pagebreak
The above can be expressed as
\begin{align}\label{app:mixed_term}
&c_1c_2 \left[\hat{a}^{\dagger2},\rho\right] +c _1^*c_2^* \left[\hat{a}^{2},\rho\right] -i\tfrac{1}{m\omega} \left[\hat{F},\hat{P} \right] \left[\left\{\hat{a},\hat{a}^\dagger\right\}, \rho\right]\nonumber \\
& - \left[\hat{F},\hat{Q} \right] \left[\{\hat{a}^\dagger\hat{a},\rho\} - 2\hat{a}^\dagger\rho\hat{a} - \left(\{\hat{a}\hat{a}^\dagger,\rho\} - 2\hat{a}\rho\hat{a}^\dagger \right) \right]\,.
\end{align}
Applying Eqs.~\eqref{app:double_com} and \eqref{app:mixed_term} to Eq.~\eqref{eq:gen_creation} we obtain
\begin{widetext}
\begin{align}
  d\rho' =& -\tfrac{i}{\hbar}[\hat{H},\rho]dt 
-  \left( \tfrac{\gamma}{16}c_1^{*2} + \tfrac{1}{4\gamma\hbar^2} c_2^{*2} \right) \left(\{\hat{a}^2,\rho\} - 2 \hat{a}\rho\hat{a}\right) dt   -  \left( \tfrac{\gamma}{16}c_1^{2} + \tfrac{1}{4\gamma\hbar^2} c_2^{2}\right)\big(\{\hat{a}^{\dagger^2},\rho\} - 2\hat{a}^\dagger\rho\hat{a}^\dagger\big) dt  \nonumber \\
&- \tfrac{i}{4\hbar}c_1c_2 \left[\hat{a}^{\dagger2},\rho\right] - \tfrac{i}{4\hbar}c _1^*c_2^* \left[\hat{a}^{2},\rho\right]dt 
- \tfrac{1}{4\hbar m\omega} \left[\hat{F},\hat{P} \right] \left[\left\{\hat{a},\hat{a}^\dagger\right\}, \rho\right]dt 
- \tfrac{1}{2} \left( c - \tfrac{i}{\hbar} \left[\hat{F},\hat{Q} \right]\right) ( \{\hat{a}^\dagger\hat{a},\rho\} - 2\hat{a}^\dagger\rho\hat{a} )  dt   \nonumber \\
&-  \tfrac{1}{2} c (\{\hat{a}\hat{a}^\dagger,\rho\} - 2\hat{a}\rho\hat{a}^\dagger )  dt  
+ \tfrac{i}{\sqrt{2\gamma}\hbar}\left[c_2^*\hat{a} + c_2\hat{a}^\dagger,\rho\right]dW 
+ \tfrac{\sqrt{\gamma}}{2\sqrt{2}}\left\{c_1^*\hat{a}+ c_1\hat{a}^\dagger - \langle c_1^*\hat{a}+ c_1\hat{a}^\dagger\rangle,\rho\right\}dW \,,
\end{align}
\end{widetext}
where 
\begin{align}
c &: =  \tfrac{\gamma}{8}|c_1|^2 + \tfrac{1}{2\gamma\hbar^2}|c_2|^2 + \tfrac{i}{2\hbar} \left[\hat{F},\hat{Q} \right] \nonumber \\
& = \tfrac{\hbar\gamma}{8m\omega} +\tfrac{1}{2\gamma\hbar m\omega} (u^2+m^2\omega^2 v^2) - \tfrac{i}{2\hbar} \left[\hat{Q},\hat{F} \right]\,.
\end{align}
In the rotating wave approximation fast oscillating terms quadratic in $\hat{a}$ or $\hat{a}^{\dagger}$ are ignored. This leads to
 \begin{align} 
   d\rho' =& -\tfrac{i}{\hbar}\left[\hat{H},\rho\right]dt  - \tfrac{1}{4\hbar m\omega}\left[\hat{F},\hat{P}\right][\left\{\hat{a},\hat{a}^\dagger\right\},\rho]dt \nonumber \\
  &+ \left(c+i\tfrac{1}{\hbar}\left[\hat{Q},\hat{F}\right] \right) (\hat{a}\rho\hat{a}^\dagger-\tfrac{1}{2}\{\hat{a}^\dagger\hat{a},\rho\})dt \nonumber \\
  &+ c (\hat{a}^\dagger\rho\hat{a}-\tfrac{1}{2}\{\hat{a}\hat{a}^\dagger,\rho\})dt\,.
\end{align}

\end{document}